\documentclass[11pt,a4paper]{article}

\newif\ifanonymized
\anonymizedfalse  % <-- Change to \anonymizedtrue for double-blind submission

\usepackage[utf8]{inputenc}
\usepackage[T1]{fontenc}
\usepackage{lmodern}
\usepackage[english]{babel}
\usepackage{geometry}
\usepackage{graphicx}
\usepackage{booktabs}
\usepackage{float}
\usepackage{hyperref}
\usepackage{amsmath}
\usepackage{cleveref}
\usepackage{enumitem}
\usepackage{xcolor}
\usepackage{authblk}
\usepackage{microtype}
\usepackage{tabularx}
\usepackage{longtable}
\usepackage{multirow}
\usepackage{xspace}

\ifanonymized
  \hypersetup{
    colorlinks=true,
    linkcolor=blue!70!black,
    citecolor=blue!70!black,
    urlcolor=blue!70!black,
    pdfauthor={},
    pdftitle={},
  }
\else
  \hypersetup{
    colorlinks=true,
    linkcolor=blue!70!black,
    citecolor=blue!70!black,
    urlcolor=blue!70!black,
    pdfauthor={William Oliveira},
    pdftitle={Throughput per Megabyte: A Pilot Benchmark of Language-Stack Efficiency for Self-Hosted HTTP Services on a Raspberry Pi 5},
  }
\fi

\newcommand{\rqone}{RQ1\xspace}
\newcommand{\rqtwo}{RQ2\xspace}
\newcommand{\rqthree}{RQ3\xspace}
\newcommand{\rqfour}{RQ4\xspace}
\newcommand{\rqfive}{RQ5\xspace}
\newcommand{\rqsix}{RQ6\xspace}

\title{Throughput per Megabyte: A Pilot Benchmark of Language-Stack\\Efficiency for Self-Hosted HTTP Services on a Raspberry~Pi~5}

\ifanonymized
  \author{[Anonymous for review]}
  \date{}
\else
  \author{William Oliveira\\
  Independent Researcher\\
  \texttt{contact@woliveiras.com}}
  \date{2026}
\fi

\begin{document}
% ============================================================

% --- Title page ---
\begin{titlepage}
  \centering
  \vspace*{\fill}
  {\LARGE\bfseries Throughput per Megabyte: A Pilot Benchmark of\\Language-Stack Efficiency for Self-Hosted HTTP Services on a Raspberry~Pi~5\par}
  \vspace{2em}
  \ifanonymized
    {\large [Anonymous for review]\par}
  \else
    {\large William Oliveira\par}
    \vspace{0.5em}
    {\normalsize Independent Researcher\par}
    {\normalsize \texttt{contact@woliveiras.com}\par}
    \vspace{1em}
    {\normalsize 2026\par}
  \fi
  \vspace*{\fill}
\end{titlepage}

% --- Abstract page ---
\begin{titlepage}
  \vspace*{\fill}
  \begin{abstract}
Cloud-centric web benchmarks miss constraints that matter for
self-hosted services on ARM64 single-board computers, especially idle
RAM footprint and energy per request.
We ran a pilot benchmark on one Raspberry~Pi~5, measuring equivalent
SQLite-backed CRUD APIs implemented in Go~1.26/net\slash http,
Rust~1.95/Axum, Python~3.13/FastAPI+Granian, Node.js~24/Fastify, and
.NET~10 Native~AOT across $N=50$ randomized runs per stack and endpoint,
with a separate concurrency sweep and memory time-series collection.
Rust achieved the highest weighted throughput-to-RAM ratio
(310.25\,req/s/MB, $2.8\times$ Go) because its weighted peak RSS was
7.36\,MB, while .NET, Node.js, Go, and Rust formed a raw-throughput
cluster with overlapping bootstrap CIs (2{,}125--2{,}461\,req/s) and
Python remained below that cluster at 969\,req/s.
The concurrency sweep on \texttt{GET /items/:id} showed divergent
scaling: .NET reached 26{,}209\,req/s at $c=240$, Go and Rust plateaued
near $c=120$, Node.js saturated near 9{,}700\,req/s, and Python stayed
below 2{,}300\,req/s; memory snapshots showed Rust holding a flat
7.8\,MB RSS while Node.js grew by 117\,MB without visible GC drops.
For RAM-constrained self-hosted CRUD-over-SQLite deployments on this
Pi~5 unit, the data support Rust as the most resource-efficient measured
stack, .NET as the strongest high-concurrency stack, and Go as a
practical option where ecosystem breadth matters; replication is needed
before extending these findings to other devices or storage backends.
  \end{abstract}
  \vspace{1em}
  \noindent\textbf{Keywords:}
  self-hosted software; ARM64; Raspberry Pi; language benchmarks;
  runtime efficiency; memory footprint; energy per request; SQLite
  \vspace*{\fill}
\end{titlepage}

% --- Table of contents ---
\tableofcontents
\thispagestyle{empty}
\newpage

% ============================================================
\section{Introduction}
\label{sec:introduction}
% ============================================================

Self-hosted software has grown rapidly, driven by privacy concerns,
cost reduction, and user control over personal data.
Projects such as Home Assistant, Jellyfin, Miniflux, Immich, n8n, and Gitea
are downloaded millions of times per year and routinely deployed on Raspberry Pi
and similar ARM64 single-board computers (SBCs)~\cite{selfhosted-reddit,homeassistant}.
These devices impose constraints that differ from server-grade hardware:
4--8--16\,GB of LPDDR4X RAM shared across all running services,
power budgets of 5--27\,W, and passive or low-RPM cooling.

The binding constraint for always-on daemon services is \emph{RAM idle footprint},
the RSS of the server process while handling zero requests.
On a Raspberry Pi~5 with 16\,GB of RAM running a home server stack
(DNS, reverse proxy, media server, automation, dashboards),
each additional 100\,MB consumed by a daemon's language stack is a real cost.
This constraint is absent from the two major public benchmarks:
TechEmpower Web Framework Benchmarks~\cite{techempower} test on server-grade x86 hardware
and report only throughput and latency; the Computer Language Benchmarks Game~\cite{clbg}
measures algorithmic CPU tasks, not HTTP servers.

Language and framework stack selection for a self-hosted service is therefore a
systems decision with long-running operational consequences, yet it lacks empirical
grounding for the ARM64/resource-constrained scenario. Practitioners rely on
informal comparisons, blog posts, and extrapolations from x86 benchmarks that do
not transfer directly.

We address this gap with a controlled \emph{pilot} experiment on a
single Raspberry~Pi~5 unit, measuring five language/framework stacks
across four dimensions directly relevant to self-hosted deployment:
RAM footprint, throughput-to-RAM ratio, startup time, and energy
consumption. All five implementations expose an identical CRUD~API
backed by SQLite, in idiomatic style for each stack.
The study additionally characterises concurrency-scaling behaviour
(throughput at $c=30$--$240$) and memory dynamics via high-frequency
time-series sampling.
The study is framed as a pilot rather than a definitive benchmark:
it uses a single hardware unit, a single embedded database (SQLite),
and a single workload family (CRUD). Energy measurements rely on the
Pi~5 onboard PMIC, whose ${\pm}10\%$ datasheet accuracy precludes
confirmatory claims among runtimes whose differences fall inside that
band.
The primary purpose is to characterise the order of magnitude of
the gaps and to motivate larger replications.
\ifanonymized
  The full replication package is available in an anonymous repository
  (link provided upon acceptance).
\else
  The full replication package, including
  benchmark scripts, raw data, and analysis code, is published at
  \url{https://github.com/woliveiras/2026-lang-self-hosted-pi} for community audit.
\fi

We investigate six research questions, all scoped to a single
Raspberry~Pi~5 unit under a SQLite-backed CRUD workload:

\begin{description}
  \item[\rqone] What is the RAM footprint (idle and under load) of equivalent HTTP API servers
    implemented in Go, Rust, Python, Node.js, and .NET on this Raspberry~Pi~5?
  \item[\rqtwo] What is the throughput-to-RAM ratio (requests per second per MB of RAM) for each
    language under this workload, and which language offers the best ratio in this setting?
  \item[\rqthree] How do startup time and binary/deployment size compare across the same stacks,
    and what are the operational implications for self-hosted services on the Pi~5?
  \item[\rqfour] What is the exploratory energy profile (average power draw and
    derived energy per request) of each language stack as reported by the
    Pi~5 onboard PMIC?
  \item[\rqfive] How does throughput scale with increasing concurrency
    ($c=30$--$240$), and where does each stack saturate on this Pi~5?
  \item[\rqsix] What memory-allocation dynamics, including heap growth,
    garbage collection, and steady-state RSS, emerge under sustained load?
\end{description}

The contributions of this paper are:
\begin{enumerate}[nosep]
  \item A pilot benchmark of five language stacks on a single
    Raspberry~Pi~5 (Cortex-A76 ARM64) unit, measuring RAM, throughput,
    startup, and energy simultaneously (\Cref{sec:results}, RQ1 and
    RQ3--RQ4). Energy results are reported as exploratory only
    (\Cref{sec:threats}).
  \item A composite metric, throughput-to-RAM ratio, operationalised
    for the self-hosted Pi~5 scenario, with a sensitivity analysis
    across three illustrative workload mixes
    (\Cref{sec:results}, RQ2).
  \item A concurrency-scaling characterisation ($c=30$--$240$) that
    reveals divergent saturation behaviour across runtimes, and a
    memory time-series analysis that identifies runtime-level memory
    management patterns (GC frequency, heap growth, steady-state
    convergence) invisible to single-point peak-RSS measurement.
  \item Practitioner-facing observations for Pi~5 deployments mapping
    deployment constraints to candidate stacks
    (\Cref{sec:discussion}, \Cref{tab:decision}). These are framed
    as observations from a pilot, not as recommendations
    generalisable to all self-hosted hardware.
  \item A complete replication package containing raw data, analysis
    scripts, server implementations, and a container image for the
    analysis pipeline (\Cref{sec:conclusion}, Data Availability).
\end{enumerate}

The word \emph{pilot} is used throughout in its statistical sense:
a small, hardware-constrained study designed to characterise effect
magnitudes and protocol feasibility, whose findings should inform
but not substitute multi-device replications.

The remainder of this paper is structured as follows.
\Cref{sec:background} reviews related work.
\Cref{sec:methodology} describes the experimental design, hardware, workload, and measurement protocol.
\Cref{sec:results} presents quantitative results for each research question.
\Cref{sec:discussion} interprets the results and provides a practitioner decision table.
\Cref{sec:threats} addresses threats to validity.
\Cref{sec:conclusion} concludes.

% ============================================================
\section{Background and Related Work}
\label{sec:background}
% ============================================================

\subsection{Self-Hosted Software}

Self-hosted software refers to applications that users deploy and operate on infrastructure
they own or control, as opposed to using a vendor's hosted service.
The practice encompasses home servers, network-attached storage, personal cloud instances,
and automation platforms.
ARM64 single-board computers have become the canonical deployment target:
the Raspberry Pi~5 (BCM2712, 4$\times$ Cortex-A76 @ 2.4\,GHz, up to 16\,GB LPDDR4X)
offers enough compute for always-on services while staying below 27\,W peak power.

\subsection{Existing Benchmarks and Their Limitations}

\Cref{tab:related-work} summarizes the studies most relevant to this work and their gaps.

\begin{table}[H]
\centering
\small
\caption{Related benchmarking studies and the gap this work fills.}
\label{tab:related-work}
\begin{tabularx}{\textwidth}{lXXXX}
\toprule
\textbf{Study} & \textbf{Hardware} & \textbf{Languages} & \textbf{Metrics} & \textbf{Gap} \\
\midrule
TechEmpower~\cite{techempower} &
  Server-grade x86 &
  300+ frameworks &
  req/s, latency &
  No ARM, no RAM idle, no energy \\
\midrule
Benchmarks Game~\cite{clbg} &
  x86 desktop &
  30+ languages &
  CPU time, memory, binary size &
  Algorithmic only; no HTTP server; no self-hosted context \\
\midrule
Pereira et al.\ 2017~\cite{pereira2017} &
  x86 &
  27 languages &
  Energy, time, memory &
  Algorithmic tasks only; no HTTP; no ARM \\
\midrule
Georgiou et al.\ 2022~\cite{georgiou2022} &
  x86 &
  12 languages &
  Energy &
  No HTTP server workload; no ARM \\
\midrule
Varghese \& Buyya 2018~\cite{varghese2018} &
  x86 server &
  Java, Python &
  req/s &
  No ARM; no RAM idle; no self-hosted framing \\
\midrule
Aroca \& Gon{\c c}alves 2012~\cite{aroca2012} &
  ARM Cortex-A9, x86 Xeon &
  N/A (single stack) &
  Power (W), throughput &
  Compares architectures (ARM vs.\ x86), not languages \\
\midrule
Georgiou et al.\ 2017~\cite{georgiou2017} &
  x86 &
  5 compiled, 5 interpreted &
  Energy, time &
  Algorithmic tasks; no HTTP; no ARM \\
\midrule
Vilhelmsson 2021~\cite{vilhelmsson2021} &
  x86 desktop &
  Node.js, Go, Apache, IIS &
  req/s, memory &
  No ARM; no energy; no self-hosted framing \\
\midrule
Berg \& Redi 2023~\cite{berg2023} &
  Raspberry Pi~400 &
  Rust (Axum) &
  req/s, latency &
  Compares protocols (REST vs.\ gRPC), not language stacks \\
\midrule
\textbf{This work} &
  Raspberry Pi~5 (ARM64) &
  5 language stacks &
  RAM, req/s, latency, startup, binary size, energy &
  Combines these dimensions on a Cortex-A76-class SBC for the self-hosted scenario \\
\bottomrule
\end{tabularx}
\end{table}

The gap addressed here is the combination of: (1)~ARM64 hardware, (2)~RAM idle measurement,
(3)~throughput-to-RAM ratio as a primary metric, and (4)~energy per request.
No published study simultaneously measures all four in the context of self-hosted HTTP services.

\subsection{Benchmarking Methodology Foundations}

Our experimental design draws on established benchmarking methodology.
Georges et al.~\cite{georges2007} demonstrated that performance studies
require rigorous statistical treatment, including multiple iterations and
confidence intervals, rather than single-run comparisons.
Kalibera and Jones~\cite{kalibera2013} proposed a systematic approach for
determining the number of repetitions needed to achieve stable measurements
in reasonable time; our 53-run protocol (50 measured + 3 warm-up) follows
their principle of quantifying measurement uncertainty.
Mytkowicz et al.~\cite{mytkowicz2009} showed that environmental factors
(link order, environment size) can introduce hidden bias, motivating our
decision to pin CPU frequency, disable background services, and reboot
between language blocks.
On the energy side, Manotas et al.~\cite{manotas2016} surveyed
practitioners' perspectives on green software engineering, confirming that
energy consumption is a growing concern but measurement tooling remains
immature, a gap we partially address with PMIC-based power logging on
commodity hardware.
Hindle~\cite{hindle2012} introduced "green mining" as a methodology for
relating software configuration to power consumption, supporting our
approach of treating energy as a first-class metric alongside throughput
and memory.

\subsection{Language Stack Selection}

The five stacks were selected to cover the dominant language choices in the self-hosted ecosystem.
\Cref{tab:language-selection} describes the selection rationale and the alternatives explicitly
rejected before data collection.

\begin{table}[ht]
\centering
\small
\caption{Language and framework selection rationale. The ``Alternative'' column documents
         rejected alternatives with justification.}
\label{tab:language-selection}
\begin{tabularx}{\textwidth}{lllX}
\toprule
\textbf{Language} & \textbf{Version} & \textbf{Framework} & \textbf{Rationale / Rejected alternative} \\
\midrule
Go & 1.26.2 & \texttt{net/http} stdlib &
  Primary hypothesis: best throughput-to-RAM ratio (H$_1$) and competitive
  raw throughput and startup time (H$_2$).
  \emph{Rejected: Fiber (fasthttp)}: faster but shifts comparison from runtime to framework. \\
\midrule
Rust & 1.95.0 & Axum 0.8.9 &
  Lower bound on resource consumption; tests whether Go's GC overhead is measurable.
  \emph{Rejected: actix-web}: historically leads TechEmpower rankings but not idiomatic for typical self-hosted services; SQLite-bound workload reduces HTTP-layer differences. \\
\midrule
Python & 3.13 & FastAPI 0.115.12 + granian 2.7.4 &
  Dominant self-hosted ecosystem (Home Assistant, Paperless-ngx). Worst-case expected baseline.
  \emph{Replaced: uvicorn}: granian is a widely adopted Rust-based ASGI server that is broadly reported as 2--3$\times$ faster than uvicorn in community benchmarks, and is a drop-in replacement. \\
\midrule
Node.js & 24 LTS & Fastify 5.8.5 &
  Dominant JS self-hosted tooling (n8n, Typebot). Single-threaded baseline.
  \emph{Rejected: hyper-express ($\mu$WS)}: C++ binding breaks the language-to-language framing. \\
\midrule
.NET & 10.0 LTS & ASP.NET Core Minimal APIs, Kestrel &
  Real self-hosted ecosystem (Jellyfin, Bitwarden, Radarr).
  Published with \texttt{PublishAot=true} (Native AOT, no JIT at runtime);
  uses \texttt{CreateSlimBuilder} and source-generated
  \texttt{System.Text.Json} serialization to satisfy the
  no-reflection constraint of Native AOT.
  \emph{Replaced: standard JIT publish}: Native AOT reduces startup time, peak RAM, and binary size; first-class .NET~10 feature; \texttt{Microsoft.Data.Sqlite} and Minimal APIs are AOT-compatible. \\
\bottomrule
\end{tabularx}
\end{table}

\clearpage

% ============================================================
\section{Methodology}
\label{sec:methodology}
% ============================================================

\subsection{Hardware}

All measurements were collected on a single Raspberry Pi~5 (\Cref{tab:hardware}).
The load generator (\texttt{wrk}) runs on a separate machine connected via
Gigabit Ethernet to the same switch as the Pi. This off-board placement
eliminates CPU contention between the load generator and the server under test,
ensuring that all four Cortex-A76 cores are available exclusively to the
benchmarked server process during measurement.

\begin{table}[ht]
\centering
\small
\caption{Benchmark hardware specifications. Firmware/bootloader/PMIC
         metadata captured on the tested unit and published in the
         replication package (\texttt{data/firmware\_info.txt}).}
\label{tab:hardware}
\begin{tabular}{ll}
\toprule
\textbf{Component} & \textbf{Specification} \\
\midrule
Board              & Raspberry Pi~5 Model~B Rev~1.1 \\
SoC                & BCM2712 (4$\times$ Cortex-A76 @ 2.4\,GHz; CPU part 0xd0b, rev r4p1) \\
RAM                & 16\,GB LPDDR4X-4267 ($\approx$34\,GB/s bandwidth) \\
Storage            & Fanxiang S500 Pro 256\,GB NVMe M.2 PCIe Gen3$\times$4 (up to 3{,}000\,MB/s) \\
NVMe adapter       & Geekworm X1001 PCIe-to-M.2 NVMe SSD Shield \\
Case               & Geekworm P579-H500 with active cooler \\
PSU                & Official Raspberry Pi~5 USB-C 27\,W (5.1\,V/5\,A PD profile) \\
USB-C cable        & UGREEN UNO PD100W (20\,V/5\,A, 5\,A e-marked, USB 2.0 data, 0.5\,m) \\
Power measurement  & Onboard PMIC (Dialog/Renesas DA9091) sampled via \texttt{vcgencmd pmic\_read\_adc} \\
OS                 & Raspberry Pi OS Lite 64-bit, Debian~13 (trixie) \\
Kernel             & 6.12.75+rpt-rpi-2712 (Debian 1:6.12.75-1+rpt1, 2026-03-11) \\
Pi firmware (\texttt{vcgencmd version}) & 2026/02/23, version \texttt{85353ce4} (release, embedded) \\
Bootloader         & commit \texttt{85353ce4...8b1331} (release), capabilities \texttt{0x0000007f} \\
Load generator     & \texttt{wrk} on a separate host, connected via Gigabit Ethernet \\
\bottomrule
\end{tabular}
\end{table}

\paragraph{System isolation.}
The following isolation measures were applied before each collection session:
(1)~CPU governor set to \texttt{performance} (\texttt{cpupower frequency-set -g performance});
(2)~swap disabled (\texttt{swapoff -a});
(3)~non-essential services stopped (bluetooth, avahi);
(4)~process list recorded before each session;
(5)~core temperature logged; runs flagged if thermal throttling detected ($>$80\,°C);
(6)~power stability verified via \texttt{vcgencmd get\_throttled} before and after each run.
A session is discarded if any undervoltage bit (\texttt{0x1}) or throttle bit
(\texttt{0x4}) becomes active during data collection. The Pi was rebooted
before each session to clear historical throttle flags
(bits \texttt{0x10000}--\texttt{0x80000}).

\subsection{Workload}

\paragraph{Workload A: CRUD API with SQLite (primary).}
All five implementations expose identical endpoints (\Cref{tab:endpoints}).
The schema is: \texttt{items(id INTEGER PRIMARY KEY, name TEXT NOT NULL, created\_at TEXT NOT NULL)}.
The database is pre-populated with 1{,}000 rows before benchmark execution.
SQLite is configured in WAL mode in all implementations.

\paragraph{Request mix.}
The benchmark exercises all five CRUD operations independently:
\texttt{GET /items} (list all), \texttt{GET /items/:id} (read one),
\texttt{POST /items} (create), \texttt{PUT /items/:id} (update),
and \texttt{DELETE /items/:id} (delete).
Each operation is benchmarked in isolation via a dedicated \texttt{wrk} Lua script,
with $N=50$ measured runs per language per endpoint.
This design allows per-operation comparison across stacks,
covering read-path (\texttt{GET}), write-path (\texttt{POST}, \texttt{PUT}),
and delete-path performance independently.
A sixth endpoint, \texttt{mixed}, exercises all five operations
in a single \texttt{wrk} session using a round-robin Lua script
that cycles through GET-list, GET-one, POST, PUT, DELETE;
it is reported separately from the per-operation analysis
and excluded from the weighted aggregate.
The \texttt{GET} and \texttt{PUT} scripts cycle through IDs 1--1{,}000
(the pre-populated rows); the \texttt{DELETE} script partitions
the ID space across \texttt{wrk} threads (stride = thread count)
so that each ID is deleted exactly once, avoiding duplicate-delete
404 responses. The database is pre-seeded with 100{,}000 rows
before each DELETE run (vs.\ 1{,}000 for other endpoints),
ensuring all requests within the 30\,s window hit real rows.
This asymmetric seed size is necessary to sustain the measured
DELETE throughput ($\approx$900--960\,req/s over 30\,s without
duplicate-delete responses); a side-effect is that DELETE runs
operate against a larger SQLite B-tree index and a different page-cache
footprint than the other endpoints. The implication is discussed in
\Cref{sec:threats}.
The database is reset between runs via \texttt{setup\_db.py}.
HTTP keep-alive is enabled (\texttt{wrk} default).

\begin{table}[ht]
\centering
\small
\caption{CRUD API endpoints. All implementations expose identical routes.}
\label{tab:endpoints}
\begin{tabular}{lll}
\toprule
\textbf{Endpoint} & \textbf{Operation} & \textbf{Description} \\
\midrule
\texttt{GET /items}        & Read all  & Return all items as JSON array \\
\texttt{GET /items/:id}    & Read one  & Return single item or 404 \\
\texttt{POST /items}       & Create    & Insert item, return created record \\
\texttt{PUT /items/:id}    & Update    & Update item name, return updated record or 404 \\
\texttt{DELETE /items/:id} & Delete    & Delete by ID, return 204 \\
\texttt{mixed}             & Round-robin & Cycles GET-list, GET-one, POST, PUT, DELETE in one session \\
\midrule
\texttt{GET /health}       & Liveness  & Returns 200 OK (used for readiness check) \\
\bottomrule
\end{tabular}
\end{table}

SQLite was chosen because self-hosted software overwhelmingly uses embedded databases
(Gitea, Miniflux, Wallabag, Vikunja all default to SQLite),
and because it avoids network I/O variability that would confound RAM and CPU measurements.

\subsection{Terminology Used in This Study}

\Cref{tab:terminology} defines the main platform, runtime, and
measurement terms used throughout the paper.

\begin{table}[ht]
\centering
\small
\caption{Terminology used in this study.}
\label{tab:terminology}
\begin{tabularx}{\textwidth}{lX}
\toprule
\textbf{Term} & \textbf{Meaning in this paper} \\
\midrule
SBC & Single-board computer, represented here by the Raspberry~Pi~5. \\
ARM64 / Cortex-A76 & The 64-bit ARM architecture and CPU core family used by the tested Pi~5. \\
CRUD & Create, read, update, and delete operations exposed as HTTP API endpoints. \\
RSS & Resident set size: physical memory mapped by the server process and reported by Linux. \\
Idle RSS & RSS after process readiness and a settle window, before benchmark traffic starts. \\
Peak RSS & Maximum RSS observed during the benchmark load window. \\
RPS & Requests per second, as reported by \texttt{wrk}. \\
p99 latency & Request latency threshold below which 99\% of measured requests completed. \\
RPS/MB & Throughput-to-RAM ratio, computed as requests per second divided by peak RSS. \\
PMIC & Power-management integrated circuit on the Pi~5, used here for exploratory power sampling. \\
Energy per request & Derived metric computed from average power, actual \texttt{wrk} duration, and total requests. \\
SQLite WAL / write lock & SQLite's write-ahead logging mode and its single-writer locking behaviour under concurrent writes. \\
Native AOT & Ahead-of-time compilation for .NET, producing a self-contained binary without runtime JIT compilation. \\
GC & Garbage collection, the runtime mechanism that reclaims unused heap allocations in managed runtimes. \\
\bottomrule
\end{tabularx}
\end{table}

\subsection{Measurements}

\Cref{tab:metrics} defines all per-run metrics.

\begin{table}[ht]
\centering
\small
\caption{Per-run measurements and derivation method.}
\label{tab:metrics}
\begin{tabularx}{\textwidth}{llX}
\toprule
\textbf{Metric} & \textbf{Unit} & \textbf{Collection method} \\
\midrule
\texttt{ram\_idle\_mb}       & MB   & RSS sampled from \texttt{/proc/<pid>/status} once, after a 2\,s settle window following readiness and before any benchmark traffic \\
\texttt{ram\_peak\_mb}       & MB   & Peak RSS from \texttt{/proc/<pid>/status} polled every 100\,ms during \texttt{wrk} window \\
\texttt{cpu\_avg\_pct}       & \%   & Average CPU usage of the server process during the \texttt{wrk} window, computed from \texttt{/proc/<pid>/stat} deltas over total \texttt{/proc/stat} jiffies; a process saturating one of four cores reads $\approx 25\%$ (not normalized per core). Collected as a diagnostic but excluded from hypothesis tests; see \Cref{sec:threats} (Construct Validity, CPU utilization ceiling). \\
\texttt{startup\_s}          & s    & Wall-clock time from process launch to first successful \texttt{GET /items} 200 OK, polled via direct socket connection every 10\,ms \\
\texttt{binary\_size\_bytes} & B    & \texttt{os.path.getsize()} of binary (Go/Rust) or deployment directory (Python/Node.js) \\
\texttt{rps}                 & req/s & \texttt{wrk} \texttt{Requests/sec} output \\
\texttt{p50\_ms}             & ms   & \texttt{wrk -L} 50th percentile latency \\
\texttt{p75\_ms}             & ms   & \texttt{wrk -L} 75th percentile latency \\
\texttt{p90\_ms}             & ms   & \texttt{wrk -L} 90th percentile latency \\
\texttt{p99\_ms}             & ms   & \texttt{wrk -L} 99th percentile latency \\
\texttt{avg\_power\_w}       & W    & Mean total SoC power during \texttt{wrk} window, computed as $\sum_i I_i \cdot V_i$ over all PMIC rails reported by \texttt{vcgencmd pmic\_read\_adc} at 1\,Hz \\
\texttt{energy\_per\_req\_mj}& mJ   & $\frac{P \times t_w}{N} \times 1000$ where $P$=avg power, $t_w$=actual \texttt{wrk} duration (as reported by \texttt{wrk}), $N$=total requests (as reported by \texttt{wrk}) \\
\midrule
\texttt{rps\_per\_mb} & req/s/MB & Derived: $\texttt{rps} / \texttt{ram\_peak\_mb}$ (primary metric for \rqtwo) \\
\bottomrule
\end{tabularx}
\end{table}

The \emph{throughput-to-RAM ratio} (\texttt{rps\_per\_mb}) operationalizes ``best balance'' as a
single measurable scalar. Higher means more efficient use of limited RAM.

\paragraph{Binary size notes.}
Binary size comparisons are not equivalent artifacts across languages.
For Go and Rust, we measure the single self-contained binary.
For Python, we measure the virtual environment directory (the realistic deployment cost).
For Node.js, we measure \texttt{node\_modules} plus the application directory.
For .NET, we measure the self-contained Native AOT publish output
(\texttt{dotnet publish -r linux-arm64 --self-contained true -p:PublishAot=true}).
This heterogeneity is discussed openly in \Cref{sec:discussion} rather than hidden.

\subsection{Experimental Protocol}

Data collection is organized in three phases executed sequentially
on the same hardware unit under identical isolation conditions.

\paragraph{Phase~1: Per-endpoint throughput (primary).}
\begin{itemize}
  \item \textbf{Runs per language:} $N = 50$ measured runs per endpoint, with 3 warmup runs discarded.
  \item \textbf{Endpoints:} Five operations benchmarked independently: \texttt{get\_list}, \texttt{get\_one}, \texttt{post}, \texttt{put}, \texttt{delete}. A sixth \texttt{mixed} workload is collected descriptively and excluded from hypothesis tests and weighted aggregates.
  \item \textbf{Languages:} Go, Rust, Python (4-worker granian), Python (1-worker granian, denoted \texttt{python-1w}), Node.js, .NET Native~AOT.
  \item \textbf{Run order:} Measured runs are randomized across languages to prevent position bias, using a deterministic pseudo-random shuffle with a fixed seed (default 42) recorded in the replication package for reproducibility.
  \item \textbf{Database reset:} \texttt{setup\_db.py} re-seeds the database before each measured run.
  \item \textbf{Cold start:} Server process killed and restarted for each run.
  \item \textbf{Thermal cooldown:} After each run, the harness blocks until CPU temperature is below the configured cooldown threshold (65\,°C by default), bounded by a 60\,s safety timeout unless overridden by \texttt{BENCH\_COOLDOWN\_MAX\_S}.
  \item \textbf{Load generation:} \texttt{wrk} runs on a separate researcher laptop used as the load-generator host, while a harness on the Raspberry~Pi starts each server and records local CPU, memory, power, and temperature measurements. For Phase~1, the effective command is \texttt{wrk -t4 -c30 -d30s -T30s -L -s <endpoint.lua> http://<pi-ip>:<port>/<path>} (4 threads, 30 concurrent connections, 30\,s duration, 30\,s socket timeout, latency histogram enabled). The Pi-side harness sends the endpoint-specific Lua script name and connection count to the load-generator host via a remote-\texttt{wrk} request file. The concurrency level was calibrated during pilot runs to avoid broad HTTP instability across the five primary endpoints; the final primary dataset contains no \texttt{wrk}-reported errors for the five main stacks on those endpoints. The chosen level therefore tests sustainable throughput rather than peak burst capacity.
  \item \textbf{Environment record:} OS version, kernel, runtime version, and compiler flags logged at
    the start of each collection session.
\end{itemize}

\paragraph{Phase~2: Concurrency sweep.}
To characterise each stack's saturation behaviour, the \texttt{GET /items/:id}
endpoint is re-tested at four concurrency levels:
$c \in \{30, 60, 120, 240\}$, with $N=50$ measured runs per level per language
(5 main languages, excluding \texttt{python-1w}).
The endpoint was chosen because it isolates CPU/runtime performance without
SQLite write-lock interference.
All other protocol parameters (thermal cooldown, cold start, randomized order)
are identical to Phase~1.

\paragraph{Phase~3: Memory time-series.}
During Phase~1 and Phase~2 runs, a concurrent monitor samples
\texttt{/proc/<pid>/\{status,smaps\_rollup,stat\}} at $\approx$130\,ms intervals,
recording per-sample RSS, PSS, VmData, and CPU utilisation.
The resulting time-series (3{,}314 snapshot files across all runs)
enables analysis of memory growth curves, garbage-collection behaviour,
and steady-state convergence beyond the single peak-RSS point
captured in earlier protocol versions.

\subsection{Statistical Analysis Plan}
\label{sec:stat-analysis}

The analysis follows a pre-specified plan executed by \texttt{analysis/analyze.py}.
Because the six endpoints span different throughput and RAM profiles
(e.g., \texttt{GET /items} serializes 1{,}000 rows while
\texttt{GET /items/:id} returns a single row),
endpoint type is a major source of within-language variance.
We therefore adopt a \emph{two-level} analysis:

\paragraph{Level~1 --- Per-endpoint analysis (primary).}
Each endpoint is analysed independently ($N=50$ per language, $k=5$ groups
for the five main languages; the auxiliary \texttt{python-1w} comparison is
reported separately in \Cref{tab:python-workers}):
\begin{enumerate}[nosep]
  \item Descriptive statistics: mean $\pm$ SD, median, IQR, 95\% CI (Student's $t$). The $t$-based CI is reported for compactness in descriptive tables; for metrics where Shapiro-Wilk rejects normality (most rate and energy metrics), the $t$-based CI is an approximation and the bootstrap CIs reported alongside the weighted aggregates (Level~2) should be regarded as the inferential reference.
  \item Shapiro-Wilk normality test.
  \item Kruskal-Wallis omnibus test with $\eta^2_H$~\cite{tomczak2014},
        computed as $\eta^2_H = (H - k + 1) / (N - k)$.
  \item Dunn's post-hoc test for pairwise comparisons. Per-metric and
        per-endpoint result tables report Bonferroni-adjusted $p$-values;
        a separate Holm-Bonferroni step-down correction is also applied
        across the full family of 400 per-endpoint pairwise tests
        (\Cref{sec:threats}).
  \item Cohen's $d$ for all pairwise comparisons.
\end{enumerate}
This level eliminates endpoint type as a confound: all 50 runs within
a group exercised the same endpoint, so observed differences reflect
language/framework effects.
The \texttt{mixed} endpoint is analysed descriptively but excluded from
hypothesis tests and weighted aggregates, as it compounds all operations
into a single measurement and would conflate per-operation effects.

\paragraph{Level~2 --- Weighted aggregate (summary).}
To produce a single representative value per language, we compute a
weighted central tendency across five per-endpoint values using a
declared \emph{workload mix} (\Cref{tab:workload-mix}).
For most metrics, the central tendency is the mean.
For energy per request, we use the \emph{median} instead of the mean
because the distribution is heavily right-skewed: energy is derived as
$P \times t / N$, so low-RPS outlier runs amplify the ratio, producing
extreme right tails. Reporting the mean would over-weight those
outliers and make the aggregate less representative of typical runs.
Bootstrap 95\% confidence intervals (10{,}000 resamples, percentile method)
are computed by resampling the 50 runs within each endpoint and
recalculating the weighted value (mean or median as appropriate) in
every iteration.

\begin{table}[H]
\centering
\small
\caption{Illustrative workload mix weights (5 endpoints).
         The ``read-heavy'' mix is used as the primary aggregate;
         ``uniform'' and ``write-heavy'' mixes are used for
         sensitivity analysis (\Cref{sec:threats}).
         Weights are not derived from empirical traffic data.}
\label{tab:workload-mix}
\begin{tabular}{llr}
\toprule
\textbf{Endpoint} & \textbf{Rationale} & \textbf{Weight} \\
\midrule
\texttt{GET /items}        & Dashboard / feed pages     & 0.30 \\
\texttt{GET /items/:id}    & Detail views, API lookups  & 0.35 \\
\texttt{POST /items}       & Content creation           & 0.15 \\
\texttt{PUT /items/:id}    & Edits                      & 0.10 \\
\texttt{DELETE /items/:id} & Cleanup, unsubscribe       & 0.10 \\
\bottomrule
\end{tabular}
\end{table}

The workload mix is a modeling assumption, not an empirical observation.
Alternative mixes are examined in the sensitivity analysis
(\Cref{sec:threats}).

\paragraph{Metrics that remain pooled.}
Three metrics are endpoint-invariant by design:
\texttt{ram\_idle\_mb} and \texttt{startup\_s} are measured once per run
before any benchmark traffic, and \texttt{binary\_size\_bytes} is a
constant per language. For these metrics, we pool all $N=300$ runs per
language (50 runs $\times$ 6 endpoints) and apply the Level~1 pipeline.

\paragraph{Correlation (\rqfour).}
Spearman $\rho$ between throughput and energy per request, computed
per language across all endpoint-level observations.

RQ4 is designated \emph{exploratory}: the PMIC instrument has limited
precision ($\pm$10\,\%, 1\,Hz sampling), and the derived
energy-per-request metric is partially tautological
($P \times t / N$). We report energy results for operational
characterization but do not draw confirmatory causal conclusions.

\subsection{Measurement Infrastructure}

The benchmark pipeline is implemented in \texttt{benchmark/benchmark.py}.
For each run, the harness: (1)~resets the database via \texttt{setup\_db.py};
(2)~launches the server process; (3)~polls \texttt{GET /items} every 10\,ms
until a 200~OK response, recording the elapsed time as \texttt{startup\_s};
(4)~samples idle RSS from \texttt{/proc/<pid>/status} after a 2\,s settle
window; (5)~invokes \texttt{wrk} on the remote load-generator host while
concurrently polling peak RSS at 100\,ms intervals, sampling PMIC power
at 1\,Hz, and recording memory time-series snapshots (RSS, PSS, VmData,
CPU) from the same 100\,ms polling loop;
(6)~parses \texttt{wrk} output, computes derived metrics, and
appends a row to the CSV; (7)~kills the server process.

Warmup iterations (3 per endpoint, language, and connection level, tagged
as \texttt{is\_warmup=true} and excluded from analysis) precede the
randomized measured runs.

Each of the five server implementations was reviewed against published
idiomatic examples and framework documentation before inclusion, to ensure
that no implementation is naively inefficient or artificially optimized.
The checklist covered: server framework idioms (routing, request parsing,
response serialization), database driver configuration (connection pooling
or reuse, SQLite WAL mode, prepared-statement handling), JSON
serialization strategy (source-generated for .NET AOT, default for the
others), and async/concurrency model (goroutines, tokio, asyncio under
granian, Node.js event loop, .NET task scheduler). The checklist and the
resulting server code are published in the replication package for
external scrutiny, since all five implementations were written by a
single author (see \Cref{sec:threats}, internal validity).

% ============================================================
\section{Results}
\label{sec:results}
% ============================================================

\subsection{RQ1: RAM Footprint}

\paragraph{Idle RAM (endpoint-invariant).}
Idle RAM is measured once per run before any benchmark traffic and is
therefore endpoint-invariant.
\Cref{tab:rq1-idle} presents idle RSS pooled across all $N=300$ runs.

\begin{table}[ht]
\centering
\small
\caption{Idle RAM (RSS, MB) before benchmark traffic.
         Mean $\pm$ SD and 95\% CI over $N=300$ runs (pooled; endpoint-invariant).}
\label{tab:rq1-idle}
\begin{tabular}{lrrrrr}
\toprule
\textbf{Language} & \textbf{Mean} & \textbf{SD} & \textbf{95\% CI} & \textbf{Median} & \textbf{IQR} \\
\midrule
Rust    &   4.97 & 0.01 & [4.97, 4.97]     &  4.97 & 0.01 \\
Go      &   9.62 & 0.11 & [9.61, 9.63]     &  9.61 & 0.13 \\
.NET    &  21.35 & 0.07 & [21.34, 21.36]   & 21.34 & 0.03 \\
Node.js &  68.11 & 0.26 & [68.08, 68.14]   & 68.06 & 0.31 \\
Python  & 216.10 & 0.42 & [216.04, 216.15] & 216.03 & 0.22 \\
\midrule
\multicolumn{5}{l}{Kruskal-Wallis: $H = 1440.80$, $p < 0.001$, $\eta^2_H = 0.96$} \\
\bottomrule
\end{tabular}
\end{table}

Idle RAM differs significantly across stacks
($H = 1440.80$, $p < 0.001$, $\eta^2_H = 0.96$),
with all pairwise Dunn comparisons significant ($p < 0.001$).

\paragraph{Peak RAM (per-endpoint).}
Peak RAM depends on endpoint workload: the \texttt{GET /items} endpoint
serializes 1{,}000 rows, inflating heap pressure, while single-row
endpoints leave the working set small.
\Cref{tab:rq1-peak} presents per-endpoint peak RAM.

\begin{table}[H]
\centering
\small
\caption{Peak RAM under load (RSS, MB) by endpoint.
         Mean over $N=50$ runs per cell.
         All five endpoints yield significant Kruskal-Wallis
         results ($H > 141$, $p < 0.001$, $\eta^2_H > 0.94$),
         with all pairwise Dunn comparisons significant ($p < 0.001$).}
\label{tab:rq1-peak}
\begin{tabular}{lrrrrr}
\toprule
\textbf{Endpoint} & \textbf{Go} & \textbf{Rust} & \textbf{.NET} & \textbf{Node.js} & \textbf{Python} \\
\midrule
\texttt{GET /items}        & 17.76 &  7.91 & 68.37 & 271.42 & 241.58 \\
\texttt{GET /items/:id}    & 19.59 &  6.62 & 37.62 &  97.84 & 234.77 \\
\texttt{POST /items}       & 22.38 &  8.11 & 39.20 & 232.09 & 236.12 \\
\texttt{PUT /items/:id}    & 19.80 &  6.79 & 38.52 & 245.15 & 235.18 \\
\texttt{DELETE /items/:id} & 21.99 &  7.88 & 39.16 &  80.24 & 233.60 \\
\midrule
Weighted (read-heavy$^\dagger$) & 19.72 &  7.36 & 47.65 & 182.90 & 236.71 \\
\bottomrule
\end{tabular}

\smallskip
\footnotesize $^\dagger$Weighted over 5 endpoints (see \Cref{sec:stat-analysis}).
\end{table}

Rust achieves the lowest footprint at both idle (4.97\,MB) and peak load
(6.62--8.11\,MB depending on endpoint), growing by only 1.6--3.1\,MB
under 30 concurrent connections.
Go occupies the second position with 9.62\,MB idle and 17.76--22.38\,MB
peak, a 8.1--12.8\,MB delta attributable to its garbage-collected heap
and goroutine stacks.
.NET~10 Native AOT shows a bimodal pattern: the \texttt{GET /items}
endpoint inflates peak RSS to 68.37\,MB (due to large response
serialization buffers), while single-row endpoints stay near 37--39\,MB.
Node.js exhibits the highest variance across endpoints: the V8 heap
reaches 232--245\,MB on \texttt{POST} and \texttt{PUT} but remains at
80--97\,MB on \texttt{GET /items/:id} and \texttt{DELETE}. This
endpoint-dependent heap growth explains the high SD previously observed
in pooled analysis.
We attribute the write-path inflation primarily to V8's
lazy garbage collection of short-lived request-body objects
(Fastify buffers each incoming JSON body before parsing, and
\texttt{node:sqlite} performs synchronous parameter binding for each
\texttt{INSERT}/\texttt{UPDATE}). Phase~3 also shows
monotonic growth of $+117$\,MB over the measurement period without
visible GC pauses, a leak-like accumulation pattern that may become
problematic under sustained long-running operation.
Python (Granian, 4 workers) peaks at 233--241\,MB across endpoints,
the highest of all stacks, with a narrow range reflecting that Python's memory
is dominated by the interpreter and imported modules rather than
per-request allocation.

For the tested Raspberry~Pi~5 with 16\,GB RAM hosting multiple co-located services,
the practical gap is large: using the weighted peak-RSS values, Rust consumes
7.36\,MB ($0.05\%$ of available RAM) while Python consumes 236.71\,MB
($1.46\%$), a $32.2\times$ per-instance memory difference.

\paragraph{Python worker-count trade-off.}
\Cref{tab:python-workers} compares the 4-worker (default granian)
and 1-worker Python configurations across all endpoints.
The single-worker variant uses $\approx$2.6$\times$ less RAM
(79--89\,MB vs.\ 216--237\,MB peak) by eliminating three worker
processes, at the cost of reduced throughput on parallelizable
endpoints.

\begin{table}[ht]
\centering
\small
\caption{Python 4-worker vs.\ 1-worker comparison (median, $N=50$ per cell).
         RAM values are peak RSS (MB); RPS/MB = throughput per MB of peak RAM.}
\label{tab:python-workers}
\begin{tabular}{lrrrrrr}
\toprule
& \multicolumn{2}{c}{\textbf{RPS}} & \multicolumn{2}{c}{\textbf{RAM peak (MB)}} & \multicolumn{2}{c}{\textbf{RPS/MB}} \\
\cmidrule(lr){2-3}\cmidrule(lr){4-5}\cmidrule(lr){6-7}
\textbf{Endpoint} & \textbf{4-w} & \textbf{1-w} & \textbf{4-w} & \textbf{1-w} & \textbf{4-w} & \textbf{1-w} \\
\midrule
\texttt{GET /items}      &   134 &    39 & 241.6 & 94.4 & 0.55 & 0.41 \\
\texttt{GET /items/:id}  & 1{,}906 & 1{,}406 & 234.8 & 89.1 & 8.19 & 15.76 \\
\texttt{POST /items}     &   612 &   615 & 236.1 & 88.5 & 2.58 & 6.94 \\
\texttt{PUT /items/:id}  & 1{,}246 &   964 & 235.2 & 90.3 & 5.30 & 10.67 \\
\texttt{DELETE /items/:id} & 614 &   625 & 233.6 & 89.3 & 2.62 & 7.00 \\
\bottomrule
\end{tabular}
\end{table}

On write-lock-bound endpoints (\texttt{POST}, \texttt{DELETE}), the
1-worker variant matches 4-worker throughput (615 vs.\ 612\,req/s)
because SQLite serializes writes regardless of process count.
On read endpoints, the 4-worker variant benefits from multi-process
parallelism: \texttt{GET /items} achieves $3.4\times$ higher RPS
(134 vs.\ 39), and \texttt{GET /items/:id} gains $1.4\times$
(1{,}906 vs.\ 1{,}406). However, when normalized by RAM consumption,
the 1-worker variant achieves \emph{higher} RPS/MB on all endpoints
except \texttt{GET /items} (where the I/O-bound serialization workload
benefits disproportionately from multi-process parallelism).
For RAM-constrained self-hosted deployments, the single-worker
configuration is therefore preferable unless the workload is
dominated by large list serializations.

\begin{figure}[ht]
  \centering
  \begin{minipage}[t]{0.48\textwidth}
    \centering
    \includegraphics[width=\textwidth]{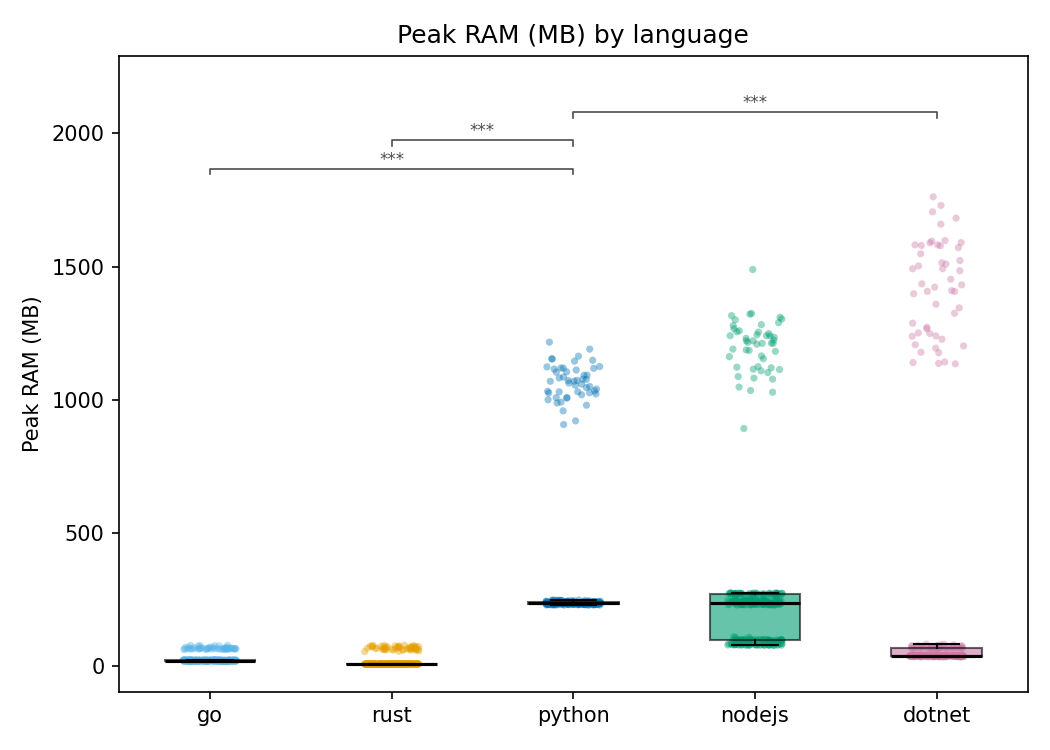}
  \end{minipage}\hfill
  \begin{minipage}[t]{0.48\textwidth}
    \centering
    \includegraphics[width=\textwidth]{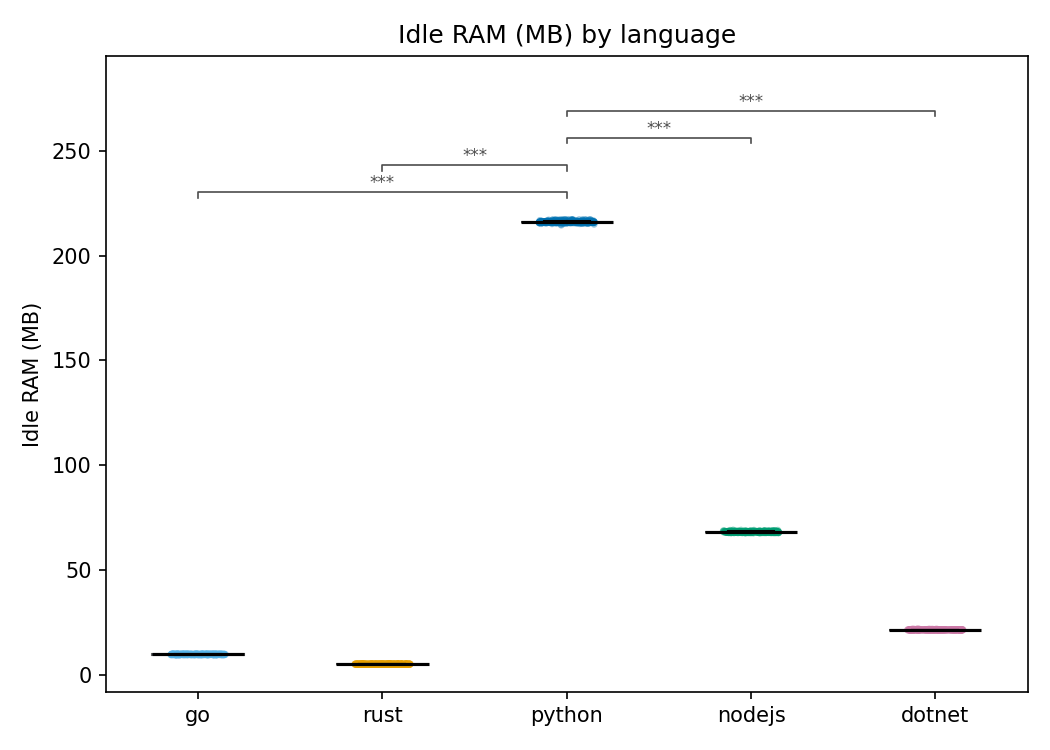}
  \end{minipage}
  \caption{Peak RAM under load (left) and idle RAM (right) by language stack.
           Box plots show median, IQR, and whiskers; individual runs overlaid
           as jittered points ($N=300$ per language, 50 per endpoint $\times$ 6).}
  \label{fig:ram}
\end{figure}

\subsection{RQ2: Throughput-to-RAM Ratio}

\paragraph{Throughput (per-endpoint).}
\Cref{tab:per-endpoint-rps} presents median throughput by endpoint.
Each cell summarizes $N=50$ runs; Kruskal-Wallis tests are applied
independently per endpoint ($k=5$ groups).

\begin{table}[ht]
\centering
\small
\caption{Median throughput (req/s) by endpoint and language stack ($N=50$ per cell).
$H$~= Kruskal-Wallis statistic; $\eta^2_H$~= effect size.
All $p < 0.001$.}
\label{tab:per-endpoint-rps}
\begin{tabular}{lrrrrrrr}
\toprule
\textbf{Endpoint} & \textbf{Go} & \textbf{Rust} & \textbf{.NET} & \textbf{Node.js} & \textbf{Python} & $H$ & $\eta^2_H$ \\
\midrule
\texttt{GET /items}      & 392  & 297  & 1{,}301 & 619  & 134  & 239.04 & 0.96 \\
\texttt{GET /items/:id}  & 4{,}128 & 4{,}049 & 4{,}229 & 4{,}186 & 1{,}906 & 159.68  & 0.64 \\
\texttt{POST /items}     & 1{,}041 & 1{,}064 & 912  & 930  & 612  & 158.68 & 0.63 \\
\texttt{PUT /items/:id}  & 3{,}957 & 3{,}702 & 3{,}970 & 3{,}788 & 1{,}246 & 197.61  & 0.79 \\
\texttt{DELETE /items/:id} & 1{,}032 & 1{,}050 & 966 & 1{,}000 & 614 & 186.90  & 0.75 \\
\bottomrule
\end{tabular}

\medskip
\footnotesize
All $p < 0.001$, all $\eta^2_H \geq 0.63$ (large to very large effects).
Dunn significant pairs (Bonferroni-corrected):
\textbf{GET /items}: .NET--all***, Go--Python***,
  Node--Python***, Rust--Python***.
\textbf{GET /:id}: all--Python***.
\textbf{POST}: Go--Python***, Rust--Python***, .NET--Python*,
  Node--Python***.
\textbf{PUT}: all--Python***.
\textbf{DELETE}: all--Python***.
(${}^{*}p<0.05$, ${}^{**}p<0.01$, ${}^{***}p<0.001$)
\end{table}

The per-endpoint analysis reveals three distinct regimes:

\begin{enumerate}[nosep]
  \item \textbf{Serialization-bound regime} (\texttt{GET /items}):
    The multi-row endpoint serializes 1{,}000 rows into JSON.
    Unlike the prior N=30 pilot (which found no language effect),
    the larger sample reveals a very large language effect
    ($\eta^2_H = 0.96$): .NET leads at 1{,}301\,req/s, likely
    due to System.Text.Json's SIMD-accelerated serialization and
    pre-compiled source generators. Go (392) and Rust (297) are
    significantly slower, suggesting their JSON libraries impose
    higher per-row overhead for large payloads.
  \item \textbf{Compute-bound regime} (\texttt{GET /:id}, \texttt{PUT}):
    Single-row read and update endpoints expose runtime overhead.
    The four compiled/JIT stacks converge (3{,}702--4{,}229\,req/s on
    \texttt{GET /:id}; 3{,}702--3{,}970 on \texttt{PUT}).
    Python is the only significantly slower stack
    (Dunn $p < 0.001$; $\eta^2_H = 0.64$--$0.79$).
  \item \textbf{Write-lock regime} (\texttt{POST}, \texttt{DELETE}):
    Write endpoints are constrained by SQLite's file-level write lock.
    Go and Rust lead ($\approx$1{,}040--1{,}064\,req/s), .NET and Node.js
    follow ($\approx$912--1{,}000\,req/s), and Python trails
    ($\approx$612--614\,req/s). The narrower spread confirms that
    the SQLite lock, not the runtime, is the throughput ceiling.
\end{enumerate}

\paragraph{Mixed endpoint (compound workload).}
A sixth endpoint exercises all five CRUD operations within a single
\texttt{wrk} script using a round-robin sequence. Under this compound
workload, sustained throughput drops to 10--25\,req/s for all stacks
(Go: 14.4, Rust: 10.6, .NET: 25.3, Node.js: 18.7, Python: 8.0\,req/s)
with peak RAM inflating to 65--1{,}427\,MB. The throughput
collapse (100--200$\times$ below single-endpoint results) reflects
SQLite's global write lock creating severe head-of-line blocking when
concurrent reads, writes, and deletes compete in the same session.
These results confirm that the per-endpoint analysis provides more
interpretable comparisons than a compound benchmark: isolating
operations separates runtime effects from database contention artifacts.
The mixed endpoint is excluded from the primary weighted aggregate
and statistical tests; it is reported here for completeness.

\paragraph{RPS/MB ratio (per-endpoint).}
\Cref{tab:per-endpoint-rpsmb} presents the throughput-to-RAM ratio
disaggregated by endpoint. All five endpoints show significant
language effects ($\eta^2_H > 0.85$) because RAM differences
amplify the ratio even when throughput differences are small.

\begin{table}[ht]
\centering
\small
\caption{Median throughput-to-RAM ratio (req/s per MB of peak RAM) by endpoint.
         $N=50$ per cell.
         All Kruskal-Wallis $p < 0.001$; all pairwise Dunn $p < 0.001$.}
\label{tab:per-endpoint-rpsmb}
\begin{tabular}{lrrrrrrr}
\toprule
\textbf{Endpoint} & \textbf{Go} & \textbf{Rust} & \textbf{.NET} & \textbf{Node.js} & \textbf{Python} & $H$ & $\eta^2_H$ \\
\midrule
\texttt{GET /items}      &  22.05 &  37.48 & 18.88 &  2.27 & 0.55  & 239.04 & 0.96 \\
\texttt{GET /items/:id}  & 210.37 & 608.64 & 112.49 & 42.67 & 8.19  & 239.04 & 0.96 \\
\texttt{POST /items}     &  46.02 & 130.54 & 23.26 &  3.99 & 2.58  & 239.04 & 0.96 \\
\texttt{PUT /items/:id}  & 199.75 & 547.35 & 103.20 & 15.46 & 5.30  & 239.04 & 0.96 \\
\texttt{DELETE /items/:id} & 46.62 & 132.53 & 24.77 & 12.40 & 2.62 & 239.04 & 0.96 \\
\midrule
Weighted (read-heavy$^\dagger$) & 111.13 & 310.25 & 60.43 & 18.93 & 4.13  & --- & --- \\
\bottomrule
\end{tabular}

\smallskip
\footnotesize $^\dagger$Weighted over 5 endpoints (see \Cref{sec:stat-analysis}).
\end{table}

Rust leads RPS/MB on every endpoint, from $1.7\times$ over Go on
\texttt{GET /items} (37.48 vs.\ 22.05) to $2.9\times$ on
\texttt{GET /items/:id} (608.64 vs.\ 210.37). This consistent advantage
stems from Rust's $2.4$--$3.4\times$ lower peak RAM across all endpoints.

\paragraph{p99 latency (per-endpoint).}
\Cref{tab:per-endpoint-p99} presents tail latency by endpoint.

\begin{table}[ht]
\centering
\small
\caption{99th-percentile latency (ms) by endpoint.
         Median over $N=50$ per cell.}
\label{tab:per-endpoint-p99}
\begin{tabular}{lrrrrrrr}
\toprule
\textbf{Endpoint} & \textbf{Go} & \textbf{Rust} & \textbf{.NET} & \textbf{Node.js} & \textbf{Python} & $H$ & $\eta^2_H$ \\
\midrule
\texttt{GET /items}        &   313 &   105 &    42 &    92 &   621 & 236.87 & 0.95 \\
\texttt{GET /items/:id}    &    12 &    12 &    11 &    11 &    29 & 114.48 & 0.45 \\
\texttt{POST /items}       &   111 &    34 & 1{,}765 &    61 & 1{,}345 & 227.24 & 0.91 \\
\texttt{PUT /items/:id}    &    16 &    16 &    16 &    40 &   497 & 181.93 & 0.73 \\
\texttt{DELETE /items/:id} &   112 &    34 & 2{,}160 &    53 & 1{,}295 & 234.13 & 0.94 \\
\bottomrule
\end{tabular}

\medskip
\footnotesize
All $p < 0.001$. Highest $\eta^2_H$ on \texttt{GET /items} (0.95)
and write endpoints (0.91--0.94).
\end{table}

Tail latency reveals a bimodal pattern for .NET: on read endpoints
(\texttt{GET /items}: 42\,ms; \texttt{GET /:id}: 11\,ms; \texttt{PUT}:
16\,ms), .NET achieves the lowest or comparable p99. On write-heavy
endpoints (\texttt{POST}: 1{,}765\,ms; \texttt{DELETE}: 2{,}160\,ms),
.NET exhibits severe tail latency spikes, suggesting periodic GC pauses
or thread-pool starvation under SQLite write-lock contention.
Python shows consistently high p99 on all endpoints (29--1{,}345\,ms).
Rust and Go show moderate and predictable tail latencies across all
endpoints.

\paragraph{Weighted aggregate summary.}
\Cref{tab:weighted-summary} presents the primary weighted aggregate
using the read-heavy workload mix (\Cref{tab:workload-mix}).
\begin{table}[ht]
\centering
\small
\caption{Weighted aggregate metrics (read-heavy mix, 5 core endpoints).
         Bootstrap 95\% CI (10{,}000 resamples).}
\label{tab:weighted-summary}
\begin{tabular}{lrrrr}
\toprule
\textbf{Language} & \textbf{RPS [95\% CI]} & \textbf{RPS/MB [95\% CI]} & \textbf{p99 (ms)} & \textbf{RAM peak (MB)} \\
\midrule
.NET    & 2{,}461 [2{,}427, 2{,}487] & 60.43  [59.57, 61.14]   & 513 & 47.65 \\
Node.js & 2{,}254 [2{,}241, 2{,}266] & 18.93  [18.72, 19.11]   &  53 & 182.90 \\
Go      & 2{,}200 [2{,}182, 2{,}214] & 111.13 [110.30, 111.85]  & 130 & 19.72 \\
Rust    & 2{,}125 [2{,}109, 2{,}140] & 310.25 [307.86, 312.40] &  48 & 7.36 \\
Python  &   969 [954, 983]           &  4.13  [4.07, 4.18]     & 590 & 236.71 \\
\bottomrule
\end{tabular}
\end{table}

Under the read-heavy mix, .NET achieves the highest weighted throughput
(2{,}461\,req/s), followed by Node.js (2{,}254), Go (2{,}200), and Rust
(2{,}125). Python trails at 969\,req/s.
The confidence intervals for the four compiled/JIT stacks overlap
substantially, confirming that throughput differences among them are
not significant in a practical sense.
The key differentiator remains RPS/MB: Rust leads at 310.25\,req/s/MB,
$2.8\times$ higher than Go (111.13) and $5.1\times$ higher than .NET
(60.43), driven by its consistently low RAM footprint across all
endpoints.

\paragraph{Effect sizes (Cohen's $d$).}
\Cref{tab:cohens-d} reports pairwise Cohen's~$d$ for four key metrics
(pooled across endpoints). RAM differences produce enormous $d$ values
($> 50$) because the distributions are tightly clustered and
well-separated. Throughput (RPS) $d$ is negligible among compiled
stacks (0.00--0.12) and medium--large only for Python pairs (0.77--0.93),
confirming the throughput cluster finding. RPS/MB $d$ values range from
medium to large across all pairs, reflecting that RAM amplifies the
efficiency ratio even when RPS differences are negligible.

\begin{table}[ht]
\centering
\small
\caption{Pairwise Cohen's~$d$ effect sizes (pooled, $N=300$ per language).
         N = negligible ($<$0.2), S = small (0.2--0.5), M = medium (0.5--0.8), L = large ($>$0.8).}
\label{tab:cohens-d}
\begin{tabular}{lrrrr}
\toprule
\textbf{Pair} & \textbf{RPS} & \textbf{RAM idle} & \textbf{RPS/MB} & \textbf{Energy/req} \\
\midrule
Go--Rust       & 0.04 (N) &  57.7 (L)  & 0.85 (L) & 0.19 (N) \\
Go--Python     & 0.80 (M) & 698.9 (L)  & 1.40 (L) & 0.42 (S) \\
Go--Node.js    & 0.00 (N) & 302.5 (L)  & 1.22 (L) & 0.09 (N) \\
Go--.NET       & 0.08 (N) & 124.4 (L)  & 0.60 (M) & 0.11 (N) \\
Rust--Python   & 0.77 (M) & 742.1 (L)  & 1.39 (L) & 0.31 (S) \\
Rust--Node.js  & 0.04 (N) & 358.3 (L)  & 1.33 (L) & 0.26 (S) \\
Rust--.NET     & 0.12 (N) & 323.1 (L)  & 1.11 (L) & 0.28 (S) \\
Python--Node.js & 0.81 (L) & 442.6 (L)  & 0.92 (L) & 0.46 (S) \\
Python--.NET   & 0.93 (L) & 674.6 (L)  & 1.40 (L) & 0.46 (S) \\
Node.js--.NET  & 0.08 (N) & 255.6 (L)  & 1.04 (L) & 0.02 (N) \\
\bottomrule
\end{tabular}
\end{table}

\begin{figure}[ht]
  \centering
  \begin{minipage}[t]{0.48\textwidth}
    \centering
    \includegraphics[width=\textwidth]{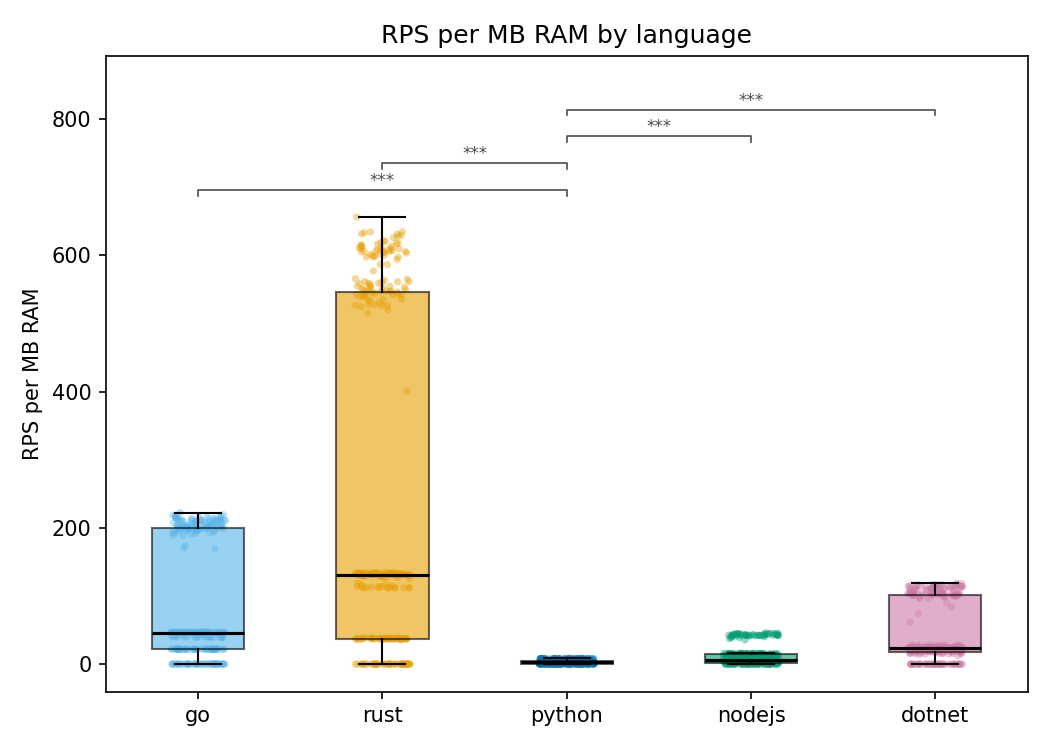}
  \end{minipage}\hfill
  \begin{minipage}[t]{0.48\textwidth}
    \centering
    \includegraphics[width=\textwidth]{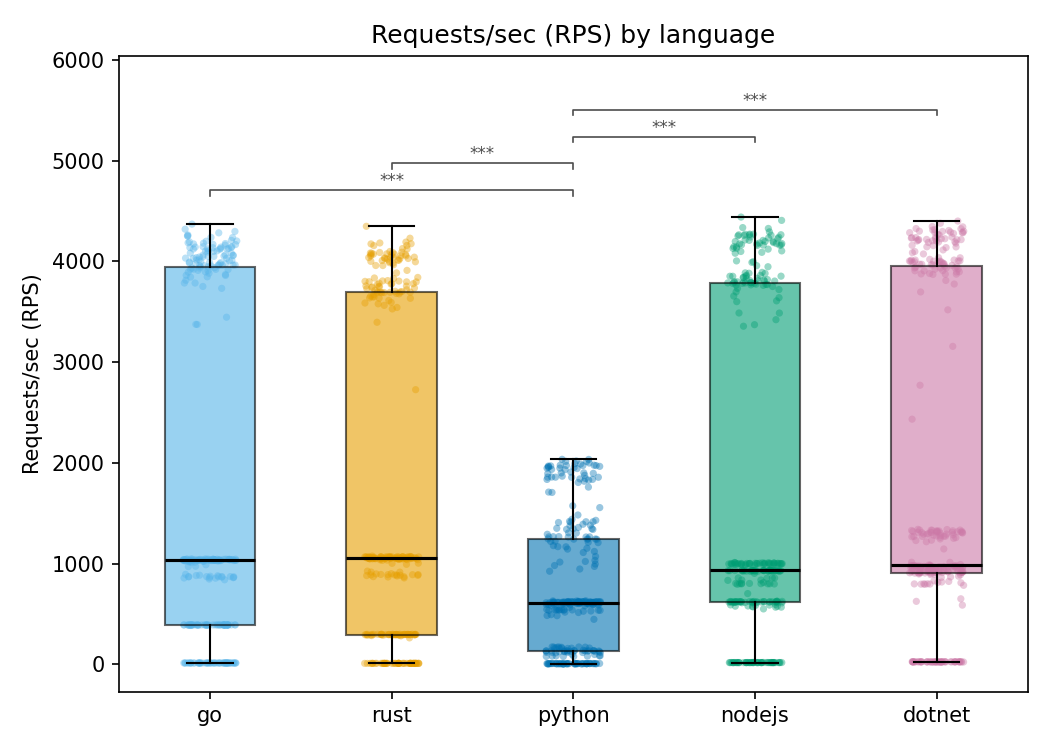}
  \end{minipage}
  \caption{Throughput-to-RAM ratio (left) and absolute throughput (right).
           Box plots with jittered individual runs ($N=300$ per language).}
  \label{fig:rps}
\end{figure}

\begin{figure}[ht]
  \centering
  \includegraphics[width=0.55\textwidth]{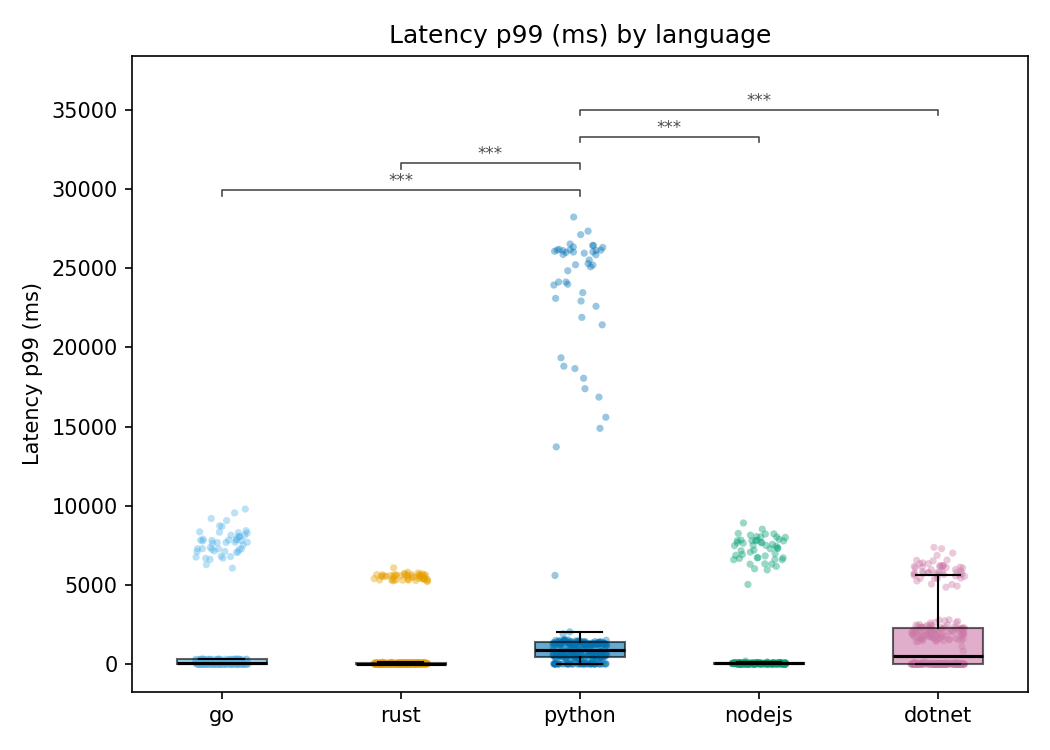}
  \caption{99th-percentile latency (ms) by language stack.
           Box plots with jittered individual runs ($N=300$ per language).}
  \label{fig:p99}
\end{figure}

\subsection{RQ3: Startup Time and Binary Size}

\begin{table}[ht]
\centering
\small
\caption{Startup time and deployment artifact size.
         $^\dagger$Entry-point script only; full deployment includes
         runtime/interpreter (see text).}
\label{tab:rq3}
\begin{tabular}{lrrrr}
\toprule
\textbf{Language} & \textbf{Startup mean (s)} & \textbf{SD} & \textbf{Artifact size} & \textbf{Total deployment} \\
\midrule
Go      & 0.0112 & 0.0003 & 8.75\,MB (static binary)  & 8.75\,MB \\
Rust    & 0.0110 & 0.0003 & 3.27\,MB (static binary)  & 3.27\,MB \\
.NET    & 0.0441 & 0.0031 & 10.19\,MB (AOT publish)   & 10.19\,MB \\
Node.js & 0.1813 & 0.0049 & 2.4\,KB (entry script)$^\dagger$  & $\approx$12\,MB \\
Python  & 0.4725 & 0.0074 & 218\,B (entry script)$^\dagger$   & $\approx$43\,MB \\
\bottomrule
\end{tabular}
\end{table}

Go and Rust start in under 12\,ms (11.2\,ms and 11.0\,ms respectively).
While the Dunn test detects a statistically significant difference
(Dunn $p < 0.001$), the 0.2\,ms gap is operationally negligible.
Both Go and Rust are effectively instant for systemd service restarts.
.NET~10 Native AOT starts
in 44.1\,ms, an order of magnitude faster than typical JIT-based
.NET deployments (which commonly exceed 1\,s~\cite{dotnet-aot}) and demonstrating
the practical benefit of ahead-of-time compilation for self-hosted
daemons. Node.js starts in 181.3\,ms and Python (Granian) in
472.5\,ms, the latter dominated by the interpreter's module
import phase.

The Kruskal-Wallis test confirms a significant omnibus difference
in startup time ($H = 1431.60$, $p < 0.001$, $\eta^2_H = 0.96$).
All pairwise Dunn comparisons are significant ($p < 0.05$).
For low-storage devices, Rust produces the smallest self-contained
binary (3.27\,MB), followed by Go (8.75\,MB) and .NET AOT
(10.19\,MB). Python and Node.js sizes in \Cref{tab:rq3} reflect
only the entry-point script; their full deployment cost includes
the interpreter/runtime (Python virtualenv $\approx$43\,MB,
\texttt{node\_modules} $\approx$12\,MB), which must be accounted
for in storage-constrained deployments.

\begin{figure}[ht]
  \centering
  \includegraphics[width=0.55\textwidth]{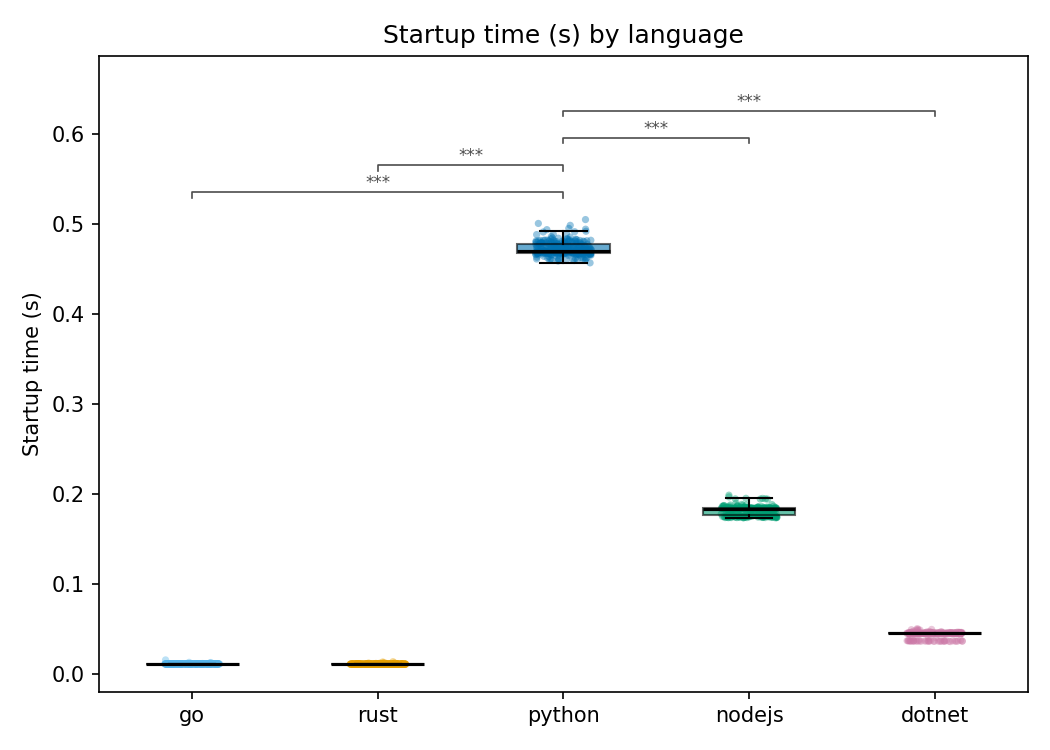}
  \caption{Startup time (s) by language stack.
           Box plots with jittered individual runs ($N=300$ per language).}
  \label{fig:startup}
\end{figure}

\subsection{RQ4: Energy Characterization (Exploratory)}

Energy metrics are endpoint-variant: faster endpoints complete more requests
for the same power draw, mechanically reducing the per-request ratio.
\Cref{tab:per-endpoint-energy} disaggregates energy per request by endpoint;
\Cref{tab:per-endpoint-power} reports the corresponding average power draw.

\begin{table}[ht]
\centering
\small
\caption{Energy per request (mJ) by endpoint.
         $N=50$ per cell; \textbf{median} reported (see \Cref{sec:stat-analysis}).
         Measured via the Pi~5 onboard PMIC (DA9091) using
         \texttt{vcgencmd pmic\_read\_adc}.}
\label{tab:per-endpoint-energy}
\begin{tabular}{lrrrrr}
\toprule
\textbf{Endpoint} & \textbf{Go} & \textbf{Rust} & \textbf{.NET} & \textbf{Node.js} & \textbf{Python} \\
\midrule
\texttt{GET /items}        & 10.89 & 15.74 &  4.87 &  7.20 & 46.68 \\
\texttt{GET /items/:id}    &  0.86 &  0.84 &  0.82 &  0.89 &  3.10 \\
\texttt{POST /items}       &  3.25 &  3.09 &  3.60 &  3.65 &  7.70 \\
\texttt{PUT /items/:id}    &  0.95 &  0.96 &  0.92 &  1.09 &  4.49 \\
\texttt{DELETE /items/:id} &  3.19 &  3.09 &  3.34 &  3.32 &  7.28 \\
\midrule
Weighted (read-heavy$^\dagger$) & 4.47 & 5.88 & 2.71 & 3.46 & 17.42 \\
\bottomrule
\end{tabular}

\smallskip
\footnotesize $^\dagger$Weighted median over 5 endpoints; bootstrap 95\% CI
(see \Cref{sec:stat-analysis}).
\end{table}

\begin{table}[ht]
\centering
\small
\caption{Average power draw during load (W) by endpoint.
         $N=50$ per cell.}
\label{tab:per-endpoint-power}
\begin{tabular}{lrrrrr}
\toprule
\textbf{Endpoint} & \textbf{Go} & \textbf{Rust} & \textbf{.NET} & \textbf{Node.js} & \textbf{Python} \\
\midrule
\texttt{GET /items}        & 4.27 & 4.65 & 6.26 & 4.44 & 6.12 \\
\texttt{GET /items/:id}    & 3.55 & 3.39 & 3.45 & 3.71 & 5.91 \\
\texttt{POST /items}       & 3.32 & 3.22 & 3.26 & 3.36 & 4.60 \\
\texttt{PUT /items/:id}    & 3.78 & 3.56 & 3.68 & 4.10 & 5.57 \\
\texttt{DELETE /items/:id} & 3.27 & 3.22 & 3.21 & 3.31 & 4.42 \\
\midrule
Weighted (read-heavy$^\dagger$) & 3.72 & 3.75 & 4.27 & 3.89 & 5.57 \\
\bottomrule
\end{tabular}

\smallskip
\footnotesize $^\dagger$Weighted over 5 endpoints (see \Cref{sec:stat-analysis}).
\end{table}

\paragraph{Per-endpoint analysis.}
Because energy per request is derived as $P \times t / N$, low-RPS
outlier runs inflate the ratio, producing heavily right-skewed
distributions. We therefore report \emph{medians}
for this metric (\Cref{tab:per-endpoint-energy}).
The \texttt{GET /items} endpoint reveals the largest energy spread:
.NET is most efficient (4.87\,mJ) due to its high throughput on this
endpoint, Go (10.89\,mJ) and Rust (15.74\,mJ) are intermediate,
and Python (46.68\,mJ) is $\approx$10$\times$ the .NET value.
On single-row endpoints, all compiled/JIT stacks cluster at
0.82--3.65\,mJ (median), while Python requires 3.10--7.70\,mJ,
a $2.0$--$3.8\times$ premium driven by its higher power draw.

\paragraph{Weighted aggregate energy.}
Under the read-heavy mix (5 core endpoints, weighted median with
bootstrap 95\% CI), the four compiled/JIT stacks fall in the
2.71--5.88\,mJ band
(.NET 2.71 [2.69, 2.74], Node.js 3.46 [3.44, 3.48],
Go 4.47 [4.44, 4.49], Rust 5.88 [5.85, 5.92]).
The full ${\pm}10\%$ PMIC accuracy on $\approx$\,3.7\,W of average
power translates to roughly ${\pm}0.5$\,mJ of uncertainty per request at
these throughput levels. The inter-stack spread within the compiled/JIT
cluster (2.71--5.88\,mJ) is larger than measurement uncertainty, but
we note that the energy ranking is driven primarily by throughput
differences (especially on the high-weight \texttt{GET /items}
endpoint) rather than power draw differences.
Python (17.42\,mJ [16.90, 17.98])
lies outside the cluster; that gap is the most defensible energy
comparison.

Average power under load varies more than in the prior pilot:
the \texttt{GET /items} endpoint drives .NET to 6.26\,W (due to
intensive JSON serialization consuming all cores) while read/write
endpoints stay at 3.2--4.1\,W for compiled stacks.
Python remains the highest at 5.57\,W weighted, a $1.4$--$1.5\times$
premium attributable to CPython's higher instruction count per request.

Within-language Spearman correlations between RPS and energy per request
are strongly negative for all five stacks ($\rho = -0.97$ to $-0.99$,
all $p < 0.001$). Because energy per request is derived as
$P \times t / N$ (power $\times$ duration $/$ total requests),
this negative relationship is partially tautological: holding power
constant, higher $N$ mechanically reduces the ratio.
We interpret energy per request as an \emph{operational efficiency
metric} rather than as evidence of an independent causal relationship.

\begin{figure}[ht]
  \centering
  \begin{minipage}[t]{0.48\textwidth}
    \centering
    \includegraphics[width=\textwidth]{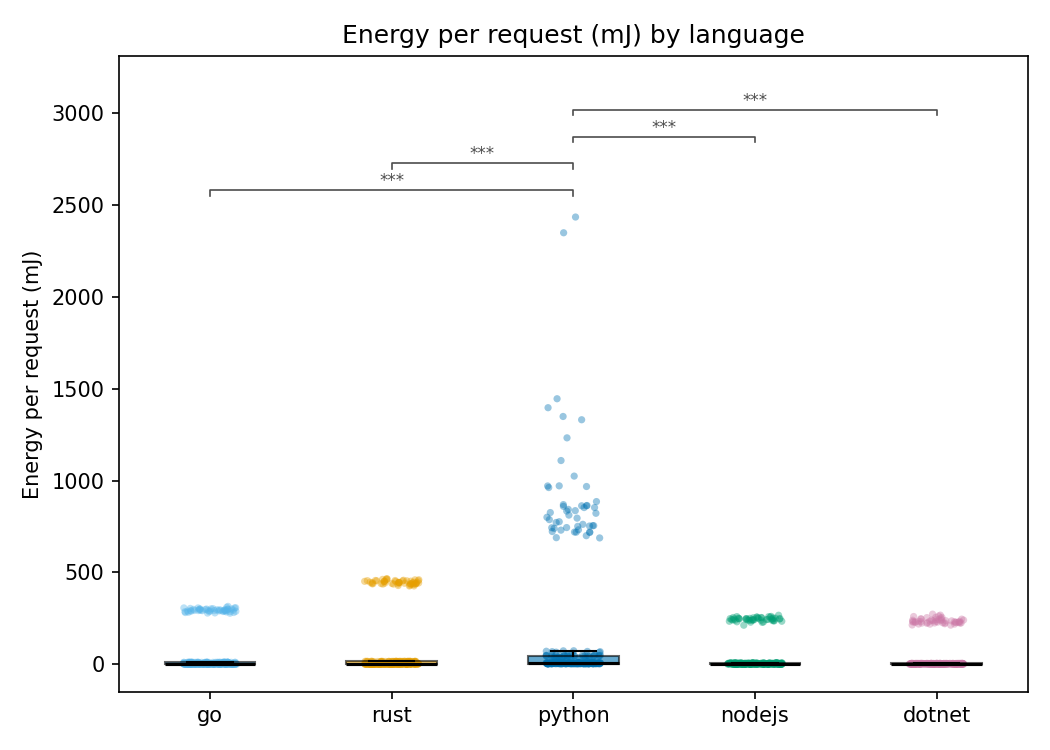}
  \end{minipage}\hfill
  \begin{minipage}[t]{0.48\textwidth}
    \centering
    \includegraphics[width=\textwidth]{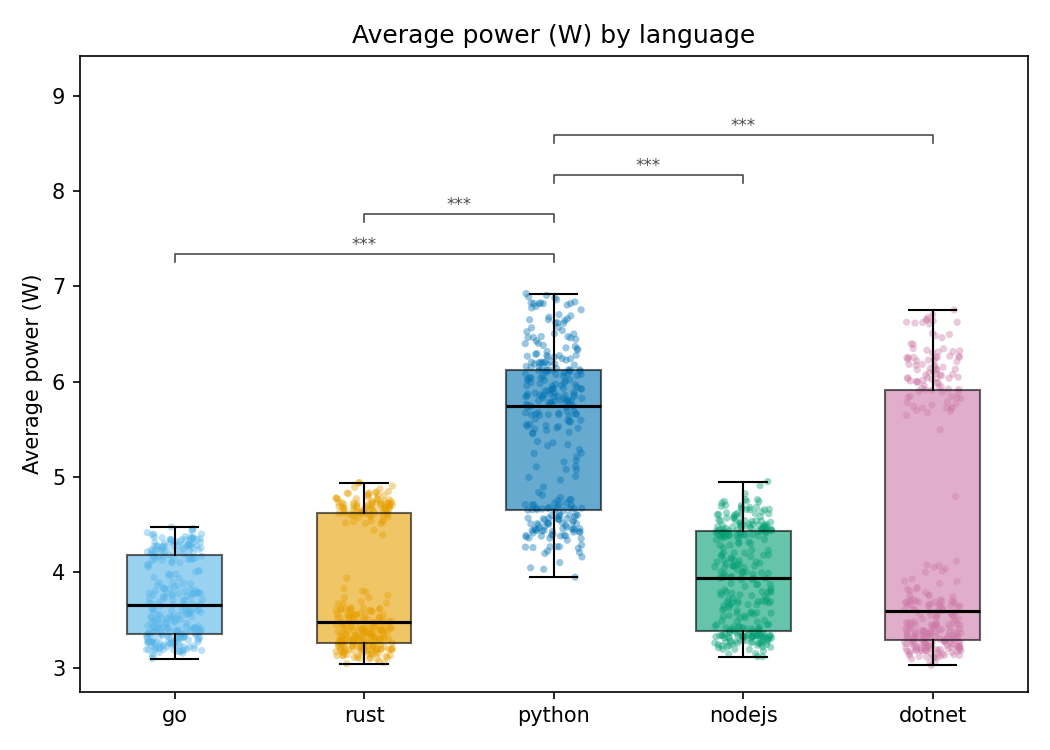}
  \end{minipage}
  \caption{Energy per request (left, mJ) and average power draw during load (right, W).
           Box plots with jittered individual runs ($N=300$ per language).}
  \label{fig:energy}
\end{figure}

\subsection{RQ5: Concurrency Saturation (Phase~2)}

Phase~2 exercises the \texttt{GET /items/:id} endpoint under increasing
concurrency ($c = 30, 60, 120, 240$) with $N=50$ measured runs per
language per level. \Cref{tab:sweep} summarizes median throughput.

\begin{table}[ht]
\centering
\small
\caption{Concurrency sweep: median throughput (req/s) on
         \texttt{GET /items/:id} at $c=30$--$240$.
         $N=50$ per cell.}
\label{tab:sweep}
\begin{tabular}{lrrrrl}
\toprule
\textbf{Language} & $c=30$ & $c=60$ & $c=120$ & $c=240$ & \textbf{Behavior} \\
\midrule
.NET    & 4{,}264 & 9{,}423 & 17{,}680 & 26{,}209 & Linear \\
Rust    & 4{,}096 & 9{,}303 & 15{,}868 & 17{,}469 & Plateau $\approx c=120$ \\
Go      & 4{,}193 & 9{,}162 & 14{,}040 & 14{,}230 & Plateau $\approx c=120$ \\
Node.js & 4{,}205 & 8{,}406 & 9{,}666  & 9{,}766  & Saturated $\approx c=120$ \\
Python  & 1{,}780 & 2{,}083 & 2{,}236  & 2{,}107  & Flat \\
\bottomrule
\end{tabular}
\end{table}

At $c=30$, the four compiled/JIT stacks cluster at
4{,}096--4{,}264\,req/s (as in Phase~1). Divergence emerges at
$c=60$ and widens sharply at $c=120$--$240$:

\begin{itemize}[nosep]
  \item \textbf{.NET} continues scaling linearly to $c=240$
    (26{,}209\,req/s, p99 = 27.3\,ms), suggesting the .NET thread pool
    and Native AOT runtime efficiently saturate all four Cortex-A76
    cores even at high connection counts.
  \item \textbf{Rust} and \textbf{Go} plateau near $c=120$
    (Rust: 15{,}868 $\rightarrow$ 17{,}469; Go: 14{,}040 $\rightarrow$ 14{,}230),
    indicating that their cooperative schedulers (tokio/goroutines)
    have saturated CPU capacity at this concurrency for this workload.
  \item \textbf{Node.js} saturates earlier, with marginal gain from
    $c=120$ to $c=240$ (9{,}666 $\rightarrow$ 9{,}766), consistent with
    the single-threaded V8 event loop being the bottleneck.
    Tail latency deteriorates sharply (p99 = 640\,ms at $c=240$).
  \item \textbf{Python} shows no meaningful scaling at any concurrency
    level (1{,}780--2{,}236\,req/s), consistent with interpreter
    overhead and the limits of process-based scaling for this workload.
\end{itemize}

The sweep reveals that Phase~1's $c=30$ protocol deliberately
underestimates the throughput headroom of .NET, Rust, and Go.
Under realistic self-hosted workloads that may reach 120--240
concurrent connections (e.g., during RSS feed refresh bursts or
reverse-proxy fan-out), .NET provides significantly higher capacity.

\begin{figure}[ht]
  \centering
  \includegraphics[width=0.65\textwidth]{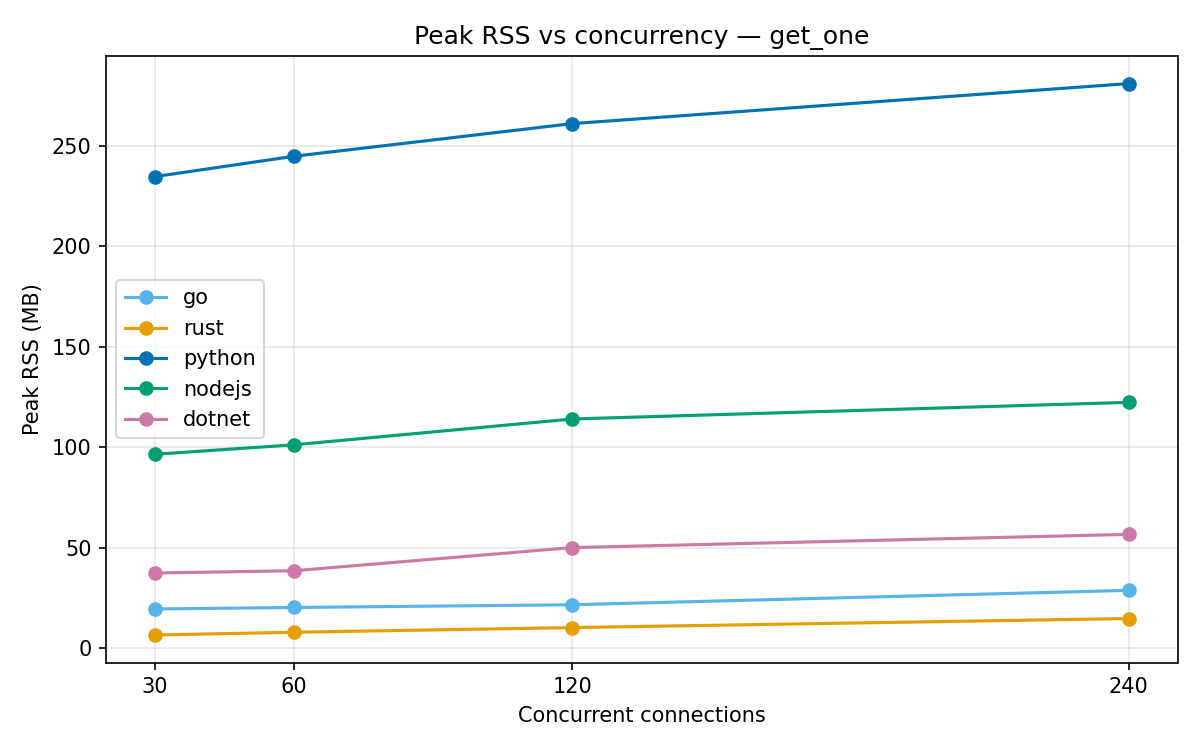}
  \caption{Peak RSS as concurrency increases from $c=30$ to $c=240$
           on \texttt{GET /items/:id}. Node.js and Python maintain
           high baselines regardless of load; .NET grows modestly
           with connection count.}
  \label{fig:sweep_rss}
\end{figure}

\subsection{RQ6: Memory Time-Series (Phase~3)}

Phase~3 captures resident set size (RSS) at $\approx$130\,ms intervals
via \texttt{/proc/[pid]/smaps\_rollup} during each benchmark run,
producing 3{,}314 time-series snapshots across all languages and
endpoints. \Cref{tab:heap} summarizes the median steady-state
characteristics.

\begin{table}[ht]
\centering
\small
\caption{Heap memory time-series summary (Phase~3, median across runs at $c=30$).
         Growth = peak $-$ initial RSS. GC drops = number of visible
         RSS decreases $>$ 0.5\,MB during a run.}
\label{tab:heap}
\begin{tabular}{lrrrrr}
\toprule
\textbf{Language} & \textbf{Peak RSS (MB)} & \textbf{Steady (MB)} & \textbf{Growth (MB)} & \textbf{GC drops} & \textbf{CPU (\%)} \\
\midrule
Rust    &   7.8 &   7.7 &   1.07 &  0 &  5.7 \\
Go      &  19.9 &  19.1 &   3.05 & 36 &  9.4 \\
.NET    &  39.1 &  34.3 &   2.75 &  4 & 10.3 \\
Node.js & 233.8 & 233.0 & 117.31 &  0 & 11.6 \\
Python  & 235.9 & 235.2 &   5.61 &  0 & 58.7 \\
\bottomrule
\end{tabular}
\end{table}

The time-series reveal qualitatively different memory management
strategies:

\begin{itemize}[nosep]
  \item \textbf{Rust}: Flat profile (7.7--7.8\,MB), negligible growth,
    no GC pauses. Memory is deterministically managed via ownership.
  \item \textbf{Go}: Active garbage collection with 36 visible
    RSS drops per measurement session. Steady-state oscillates between
    $\approx$17--20\,MB with periodic compaction.
  \item \textbf{.NET}: Moderate footprint (34--39\,MB) with 4 observed
    GC collections. The Native AOT runtime retains the .NET GC
    but triggers it less frequently than Go's collector.
  \item \textbf{Node.js}: Monotonically increasing RSS with no
    observed GC cycles (growth = $+117$\,MB over the measurement
    window). This pattern suggests either V8's old-generation GC
    threshold is not reached within a single run, or deferred
    allocations accumulate without collection.
    Long-running deployment implications are significant: absent
    explicit memory limits, Node.js may continue increasing RSS.
  \item \textbf{Python}: High baseline (235\,MB) with moderate growth
    (+5.6\,MB). No GC drops are visible in RSS (CPython uses
    reference counting with a cycle collector that does not return
    pages to the OS). CPU utilisation is highest (58.7\%) despite
    lowest throughput, confirming interpreter overhead.
\end{itemize}

\begin{figure}[ht]
  \centering
  \includegraphics[width=0.75\textwidth]{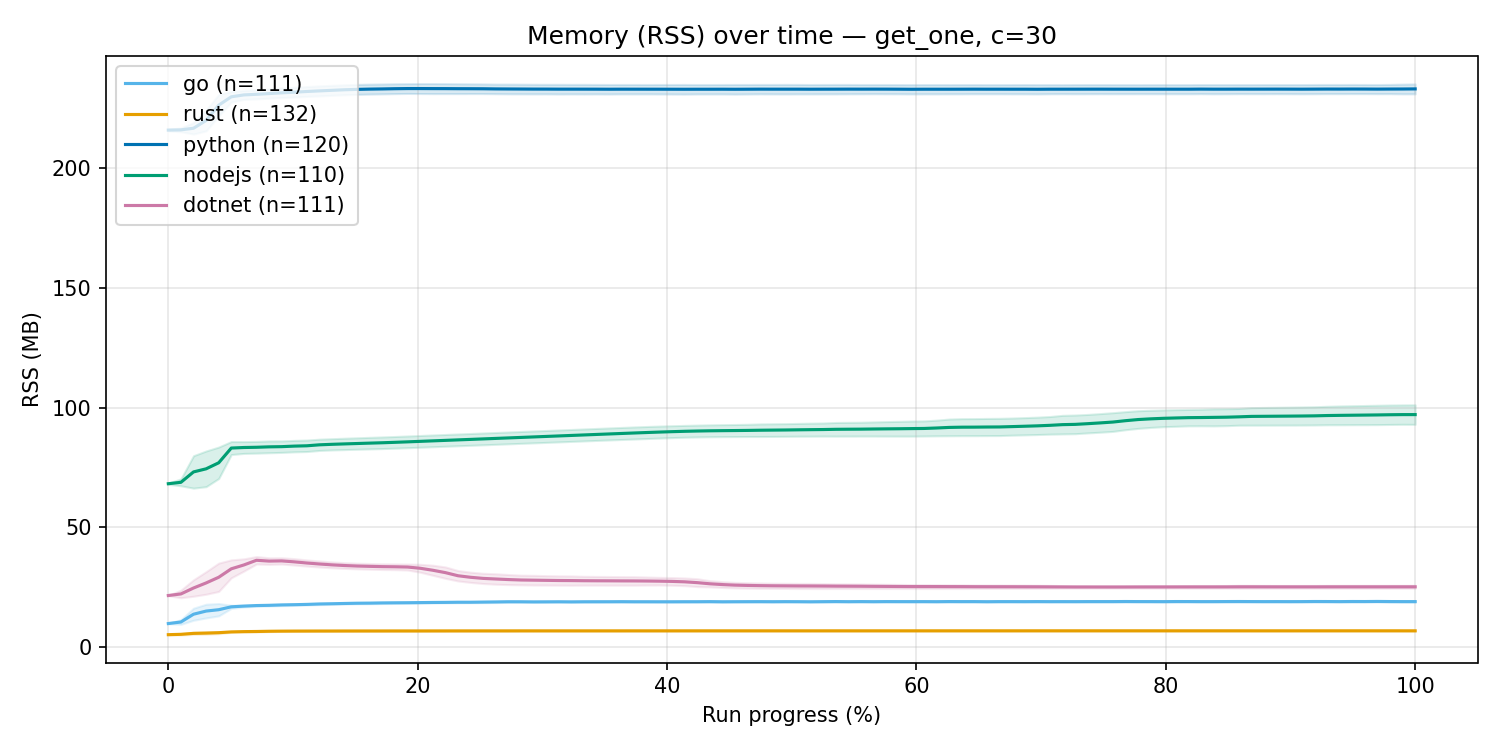}
  \caption{RSS time-series for \texttt{GET /items/:id} at $c=30$
           (representative run per language). Go exhibits visible GC
           sawtooth; Node.js grows monotonically; Rust stays flat.}
  \label{fig:memory_curve}
\end{figure}

\clearpage

% ============================================================
\section{Discussion}
\label{sec:discussion}
% ============================================================

\subsection{The Efficiency Frontier}

\begin{figure}[ht]
  \centering
  \includegraphics[width=0.65\textwidth]{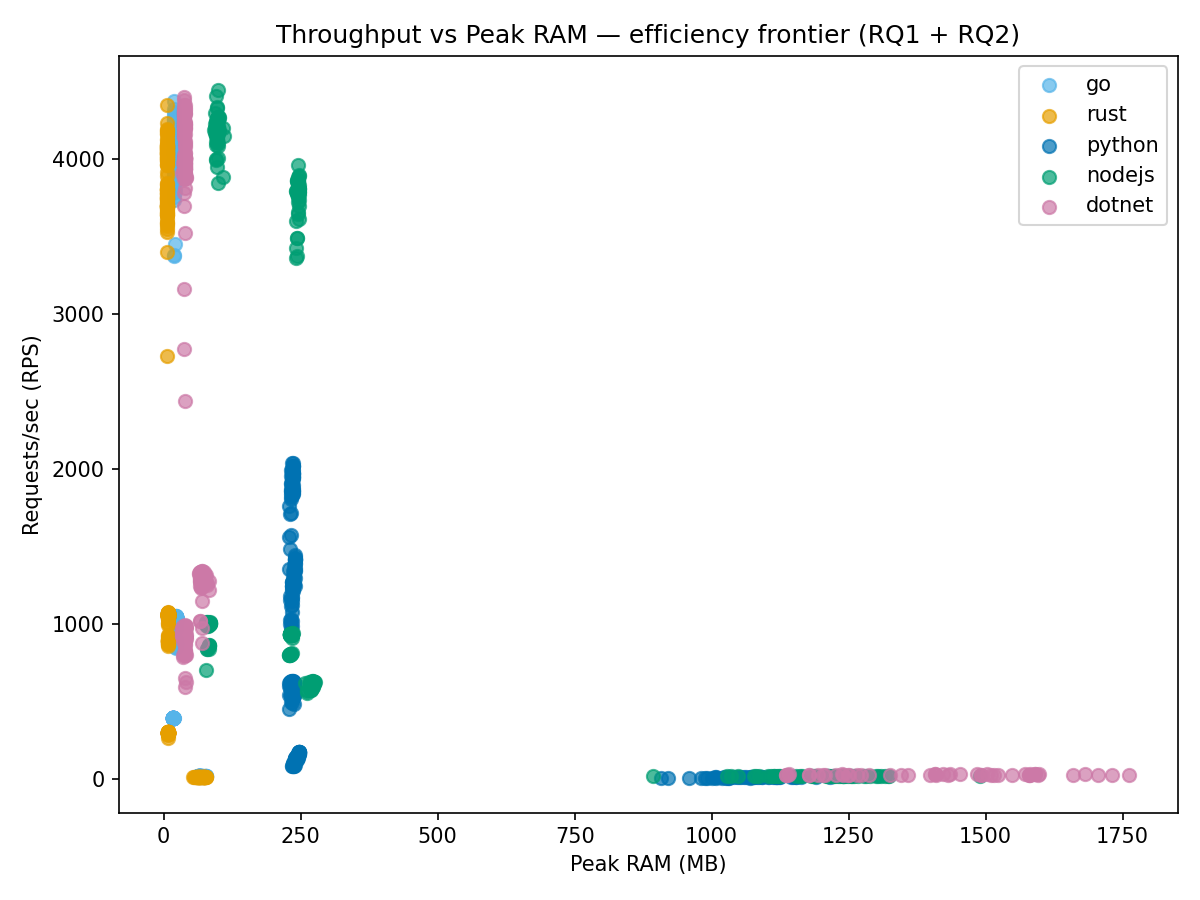}
  \caption{Throughput (RPS) versus peak RAM (MB) per language stack.
           Each point is a single run ($N=300$ per language).
           Rust occupies the upper-left Pareto corner.}
  \label{fig:scatter}
\end{figure}

The scatter plot of throughput versus peak RAM (\Cref{fig:scatter}) reveals
a clear Pareto frontier.
Rust occupies the upper-left corner: competitive weighted throughput
(2{,}125\,req/s) at the lowest weighted peak RAM (7.36\,MB), making it
Pareto-optimal when both metrics matter.
Go lies adjacent, with
similar throughput (2{,}200\,req/s; overlapping CIs) but at
$2.7\times$ the RAM cost (19.72\,MB).
No other language exceeds either Rust or Go on both axes simultaneously.

.NET~10 Native AOT achieves the highest weighted throughput
(2{,}461\,req/s) at a moderate RAM cost (47.65\,MB).
Node.js achieves competitive throughput (2{,}254\,req/s) but is
dominated by Go on RAM by a factor of $9.3\times$ and by Rust by a
factor of $24.9\times$.
Python is Pareto-dominated on both axes: lowest throughput (969\,req/s) at
the highest RAM of all stacks (236.71\,MB).

For self-hosted CRUD-over-SQLite workloads on this Raspberry~Pi~5 unit,
the Pareto frontier observed in our pilot reduces to two practical
candidates:
Rust when RAM is the binding constraint, and Go when developer
ergonomics and ecosystem breadth outweigh the RAM premium.
Whether this frontier holds on other Cortex-A76 boards, with
non-SQLite storage backends, or at higher concurrency is an open
empirical question (\Cref{sec:threats}).

\subsection{Practical Implications: Observations for Pi~5 Deployments}

\Cref{tab:decision} summarises observations from this pilot for common
self-hosted deployment scenarios on a Raspberry~Pi~5.
The table reports candidate stacks under each constraint based on the
measured data; it is not a generalised recommendation table and should
not be transferred to other hardware or workloads without replication.

\begin{table}[H]
\centering
\small
\caption{Candidate stack per self-hosted deployment scenario on the
         tested Raspberry~Pi~5 unit, under a single-process
         CRUD-over-SQLite workload.
         Rows marked $\dagger$ reflect qualitative practitioner
         judgement, not a metric measured in this study. The energy
         row is exploratory only: differences among the four
         compiled/JIT stacks fall within PMIC measurement
         uncertainty (\Cref{sec:threats}).}
\label{tab:decision}
\begin{tabularx}{\textwidth}{>{\raggedright\arraybackslash}p{0.30\textwidth}
                            >{\raggedright\arraybackslash}p{0.22\textwidth}
                            >{\raggedright\arraybackslash}X}
\toprule
\textbf{Scenario} & \textbf{Candidate stack} & \textbf{Basis} \\
\midrule
Always-on daemon with co-located services &
  Rust &
  Lowest idle footprint: 4.97\,MB (empirical). \\
High-throughput API (single service) &
  Rust &
  Highest weighted RPS/MB: 310.25\,req/s/MB (empirical). \\
Energy-constrained / battery-backed &
  Compiled/JIT cluster &
  Exploratory only: .NET, Go, Rust, and Node.js fall within 2.71--5.88\,mJ; Python is higher at 17.42\,mJ. \\
High-concurrency single endpoint &
  .NET Native AOT &
  Phase~2: 26{,}209\,req/s at $c=240$ on \texttt{GET /:id}, with linear scaling. \\
Rapid deployment / scripted service &
  Go$^\dagger$ &
  Qualitative judgement based on ergonomics and ecosystem, not a measured metric. \\
\bottomrule
\end{tabularx}
\end{table}

\subsection{Go Hypothesis Evaluation}

The primary hypothesis comprised two testable claims:
\begin{description}
  \item[H$_1$] Go has the best throughput-to-RAM ratio among the selected stacks.
  \item[H$_2$] Go has competitive raw throughput and startup time.
\end{description}

\noindent
(These hypotheses were narrowed during study design from a broader
initial expectation that Go would dominate all resource-efficiency
metrics. The split isolates the testable ratio claim from the
weaker competitiveness claim.)

\paragraph{H$_1$ (refuted).}
Rust exceeds Go on RAM footprint and throughput-to-RAM ratio
($2.8\times$ better weighted RPS/MB: 310.25 vs.\ 111.13;
$2.7\times$ lower weighted peak RAM: 7.36 vs.\ 19.72\,MB).
Rust also produces a smaller binary (3.27 vs.\ 8.75\,MB).
H$_1$ is therefore refuted: Rust, not Go, achieves the best
throughput-to-RAM ratio.

\paragraph{H$_2$ (supported).}
Under the read-heavy workload mix, the four compiled/JIT stacks
achieve overlapping throughput CIs (2{,}125--2{,}461\,req/s),
confirming no practical throughput advantage for any one stack.
Only Python trails significantly (969\,req/s).
Go achieves marginally faster startup (11.1\,ms vs.\ 11.0\,ms;
Dunn $p < 0.001$, significant but operationally negligible).
H$_2$ is supported: Go is throughput- and startup-competitive
with the other compiled/JIT stacks.

\paragraph{Practical interpretation (not measured).}
Go's broader ecosystem for HTTP middleware, its gentler learning curve,
and its extensive use in existing self-hosted projects (Gitea, Syncthing,
Caddy) are qualitative factors not measured in this study.
Whether they outweigh Rust's quantitative efficiency advantage on the
Pi~5 depends on the practitioner's priorities. Within the resource
envelope tested here, the data are consistent with Rust offering the
best measured resource efficiency for purely resource-constrained
deployments.

\subsection{Native AOT .NET}

.NET~10 Native AOT eliminates JIT startup overhead: at 44.1\,ms, its startup
time is $4.1\times$ faster than Node.js (181.3\,ms) and $10.7\times$ faster
than Python (472.5\,ms), placing it in the same order of magnitude as the
compiled-systems languages.
Idle RAM (21.35\,MB) and weighted peak RAM (47.65\,MB) are substantially lower
than what a standard JIT-based .NET publish would exhibit (typically
80--150\,MB for equivalent workloads), confirming that Native AOT
makes .NET viable for constrained Cortex-A76-class ARM64 deployment.

.NET achieves the highest weighted throughput (2{,}461\,req/s under the
read-heavy mix), with its CI overlapping the other compiled/JIT stacks.
Per-endpoint analysis reveals a bifurcated latency profile:
.NET's p99 latency spikes on the two endpoints that trigger SQLite
write locks combined with response serialization
(\texttt{POST} = 1{,}765\,ms, \texttt{DELETE} = 2{,}160\,ms;
\Cref{tab:per-endpoint-p99}), while staying within 11--42\,ms on
read and update endpoints.
Three non-exclusive mechanisms are consistent with this pattern:
(i) tail interactions between the .NET thread pool and SQLite's
file-level write lock under bursty arrivals;
(ii) GC-equivalent compaction pauses in the Native AOT runtime
(Native AOT still uses the .NET GC, only the JIT is removed);
and (iii) System.Text.Json source-generator overhead on small
write payloads. Pinpointing the dominant factor requires
\texttt{dotnet-trace} or \texttt{perf} flame-graph profiling under
the same workload, which is left for future work.
A notable finding is that .NET draws comparable weighted average power
to Go and Rust (4.27, 3.72, and 3.75\,W respectively),
confirming that Native AOT produces efficient CPU utilization.

\subsection{Python Worker Configuration Trade-off}

The 4-worker vs.\ 1-worker Python comparison
(\Cref{tab:python-workers}) illuminates a fundamental deployment
trade-off: multi-process parallelism purchases throughput on read
endpoints at the cost of proportionally higher RAM.
The 4-worker configuration spawns four OS processes, each loading
the full Python interpreter and application modules, yielding
$\approx$2.6$\times$ higher peak RAM (234--242\,MB vs.\ 89--94\,MB).
On write-bound endpoints (POST, DELETE), throughput is identical
because SQLite serializes writes regardless of process count.
On read endpoints, the benefit ranges from $1.4\times$ (GET /:id)
to $3.4\times$ (GET /items), the latter reflecting that
serializing 1{,}000 rows benefits disproportionately from
distributing rows across workers.

Critically, when normalized by RAM consumption, the single-worker
variant achieves \emph{higher} RPS/MB efficiency on 4 of 5 endpoints.
For a Raspberry~Pi~5 hosting multiple co-located services under
memory pressure, this means the 1-worker configuration provides
better utilization of the scarce RAM resource. The only exception
is GET /items (large-payload serialization), where the parallelism
gain outweighs the RAM penalty in the efficiency ratio.

This finding generalizes beyond Python: any language runtime that
supports multi-process scaling (e.g., Node.js cluster mode,
Go with GOMAXPROCS tuning) faces the same trade-off on
memory-constrained hardware. The data suggest that right-sizing
worker count to available RAM is a first-order optimization for
self-hosted deployments, not merely a configuration detail.

\subsection{Energy Implications at Fleet Scale}

The per-request energy differences are small in absolute terms
(2.71--17.42\,mJ weighted median) but compound over sustained operation.
The measured $\approx$6.4$\times$ gap between Python and .NET,
applied to a continuously busy device, accumulates into kilowatt-hour-scale
differences per device per year, and into proportionally larger differences
across a fleet of similar devices.

We deliberately stop short of converting this gap into specific
fleet-scale kilowatt-hour or tonne-CO\textsubscript{2} estimates.
Three factors make any such number unreliable for self-hosted contexts:
(i) real self-hosted services have intermittent traffic, not sustained
100\,req/s, so the relevant integration window is dominated by idle
states rather than by the per-request term;
(ii) idle board power (2.5--3\,W for the Pi~5 at rest) is the same
across language stacks and dominates total consumption during low-traffic
periods;
(iii) the grid carbon intensity that would convert kilowatt-hours into
CO\textsubscript{2} varies by over $40\times$ across jurisdictions
(e.g.\ Norway vs.\ coal-heavy regions), so the carbon impact depends more
on \emph{where} a device is deployed than on \emph{what language} it runs.
The ${\pm}10\%$ PMIC accuracy further widens the uncertainty band on
any extrapolated total.

The defensible operational statement is therefore qualitative: the
measured energy gap between Python and the compiled/JIT cluster does
not vanish at scale, and language-stack choice is a lever worth
considering alongside hardware selection and grid decarbonization for
practitioners operating energy-constrained or battery-backed fleets.
Quantitative fleet-scale carbon accounting requires a study designed
around realistic traffic traces and a calibrated power profiler
(\Cref{sec:conclusion}, future work).

% ============================================================
\section{Threats to Validity}
\label{sec:threats}
% ============================================================

\subsection{Internal Validity}

\paragraph{Thermal interference.}
The Raspberry Pi~5 may thermal-throttle under sustained load.
We mitigated this by: (1) enforcing a cooldown wait between runs
(temperature $\leq$ baseline $+$ 5\,°C), and (2) flagging any run
where throttling was detected. Throttled runs are excluded from analysis.

\paragraph{Run order bias.}
Systematic run ordering could confound results if, for example, thermal state
accumulates across languages. We randomize run order across languages per repetition.

\paragraph{SQLite write-lock ceiling.}
SQLite uses file-level locking: concurrent writers serialise behind
a single lock, regardless of how many goroutines, tokio tasks, or
worker threads the server runtime can schedule. Under our 30
concurrent connections, write-path endpoints (\texttt{POST},
\texttt{PUT}, \texttt{DELETE}) hit this ceiling, which directly
limits the throughput differentiation between language stacks on
those endpoints. The observed convergence of the four compiled/JIT
stacks on write-path throughput (\Cref{tab:per-endpoint-rps}) is
therefore not purely a language-runtime finding: a non-trivial
portion of it reflects SQLite's locking discipline, which is
identical across all stacks. The same caveat applies to the
throughput-to-RAM ratio on write endpoints. We retain SQLite
because it is the dominant storage backend for the self-hosted
applications motivating this study (Gitea, Miniflux, Vikunja,
Wallabag), but we note that a replication with a client-server
database (e.g.\ PostgreSQL or MariaDB) would lift the write-lock
ceiling and is required before any claim about pure language-runtime
throughput differences on write paths can be made. Read-path
endpoints (\texttt{GET /items/:id}, where the multi-row I/O cost
is low) are less affected by this ceiling and provide a more
direct runtime comparison.

\paragraph{Single operator implementation.}
All five server implementations were written by the same person, which may
introduce unintentional bias toward or against certain languages.
This is the dominant internal-validity threat for any
language-comparison benchmark: subtle implementation choices (buffer
sizes, connection-pooling strategy, serialization configuration, async
idioms) can shift per-endpoint results by tens of percent.
We applied three mitigations.
First, each implementation was reviewed against published idiomatic
examples and framework documentation before inclusion using a
standardized checklist (framework idioms, database driver
configuration, JSON serialization, concurrency model) recorded in the
replication package.
Second, all five server implementations, the checklist, and the raw
data are published openly to enable external review and the
submission of language-specific patches.
Third, where per-endpoint behaviour is suspicious (notably the Node.js
write-path heap inflation on \texttt{POST}/\texttt{PUT};
\Cref{sec:results}), the paper discusses the mechanism explicitly
rather than reporting the number in isolation.
We nonetheless caution that residual implementation bias cannot be
fully ruled out without independent re-implementation by domain
experts in each language, which is left for future work.

\paragraph{Discarded failed runs.}
Runs in which the load generator (\texttt{wrk}) returns a non-zero exit code,
times out, or produces output that the harness cannot parse are not written
to the dataset. This prevents contamination of the analysis with rows containing
zeros for \texttt{rps} and latency. Such failures are logged with the captured
server stderr for post-hoc diagnosis.
Across all five stacks, no measured runs required discarding during the
final data-collection session; every stack produced $N=50$ valid runs
per endpoint as planned.

\paragraph{Network-based load generator.}
\texttt{wrk} runs on a separate machine connected to the Raspberry~Pi~5 via
Gigabit Ethernet through the same switch. This eliminates CPU contention between
the load generator and the server under test, allowing all four Cortex-A76 cores
to serve the benchmarked process exclusively. However, it introduces a network
hop that adds a small amount of latency jitter. We mitigate this by using a
direct Ethernet connection (no WiFi, no router hops) and by reporting latency
percentiles rather than averages, which are less sensitive to occasional
network outliers. Absolute latency values should be interpreted with this context; the
relative ranking between stacks remains valid since all five stacks traverse
the same network path.

\paragraph{Page cache.}
The Linux page cache is not explicitly flushed (\texttt{drop\_caches})
between runs. Cached SQLite database pages from a previous run could
artificially benefit the next run's I/O path. Because all five stacks
use the same database file, the same OS, and the same storage device,
the caching advantage is symmetric across languages for steady-state
throughput, and the 30-second \texttt{wrk} window ensures that the
steady state (not cold-cache I/O) dominates each measurement.
The argument is weaker for \texttt{startup\_s}: the first run
immediately after a session-start reboot incurs cold-cache I/O for
shared libraries, while subsequent runs benefit from cached pages.
We mitigate this in two ways. Run order is randomized across
languages, so any first-run-after-reboot effect is distributed across
stacks rather than concentrated on one of them, and three warm-up
runs per stack per endpoint are discarded before measurement begins.
Adding \texttt{echo 3 > /proc/sys/vm/drop\_caches} before each run
would further isolate the startup-time measurement and is a
straightforward extension for replications.

\paragraph{Heap-snapshot sampling overhead.}
Phase~3 memory time-series collection reads
\path{/proc/[pid]/smaps_rollup}
at $\approx$130\,ms intervals from an external monitoring script.
Although \texttt{smaps\_rollup} is a lightweight kernel interface (no
\texttt{ptrace} attach), each read briefly holds the target process's
\texttt{mmap\_lock} for reading, which may introduce sub-microsecond
pauses. We consider this overhead negligible relative to the 30\,s
measurement window, but cannot completely rule out minor perturbation
of GC timing or allocation patterns for managed runtimes.

\subsection{External Validity}

\paragraph{Single hardware unit.}
All measurements were collected on a single Raspberry~Pi~5 unit.
Unit-to-unit manufacturing variance (silicon binning, thermal-paste
application, PSU + cable tolerance, ambient airflow inside the case)
can affect both absolute throughput and absolute power, and the
magnitude of that variance is not characterised by this study.
This is the dominant external-validity threat and the primary reason
the paper is framed as a pilot rather than as a definitive benchmark.
Relative rankings between stacks that differ by an order of magnitude
on a given metric (e.g.\ idle RAM, RPS/MB) are likely to be stable
to unit-to-unit variance; close comparisons (e.g.\ Go vs.\ Rust
startup at sub-millisecond difference, or the four-way energy cluster
inside the PMIC's $\pm 10\%$ band) are not.
Replication on at least three additional Pi~5 units, and ideally on
other Cortex-A76-class boards (Pi~4 update, Orange~Pi~5, Rock~5),
is required before any of the close comparisons can be treated as
established. Results may also differ on boards with different memory
subsystems, microarchitectures, or thermal envelopes.

\paragraph{Workload representativeness.}
The benchmark exercises all five CRUD operations (\texttt{GET} list, \texttt{GET} one,
\texttt{POST}, \texttt{PUT}, \texttt{DELETE}) independently plus a sixth
\texttt{mixed} endpoint that cycles through the operations in a
round-robin \texttt{wrk} script.
Independent endpoints isolate per-operation performance. The mixed
endpoint is not used to validate the weighted aggregate because compound
traffic introduces head-of-line blocking through SQLite's write lock and
therefore measures database contention as much as runtime behaviour.
The \texttt{DELETE} endpoint is inherently destructive: each request
removes a row from the database. To avoid 404 contamination from
duplicate deletes, the database is pre-seeded with 100{,}000 rows
before each DELETE run, and the \texttt{wrk} Lua script partitions
the ID space across threads (stride = thread count) so that each
ID is targeted by exactly one thread.
At $\approx$\,900--1{,}050\,req/s $\times$ 30\,s $\approx$ 27{,}000--31{,}500
requests per run, the 100{,}000-row seed provides ample headroom.
In the final primary dataset, the measured DELETE rows contain no
\texttt{wrk}-reported errors after this 100{,}000-row seeding and
ID-partitioning change.
Note that the 100{,}000-row seed means DELETE operates against a
larger SQLite index than the 1{,}000-row seed used for other
endpoints; this may introduce a minor index-lookup asymmetry,
though SQLite's B-tree lookup remains $O(\log n)$ in both cases.

\paragraph{Framework versions.}
Framework versions are fixed at the time of data collection (\Cref{tab:language-selection}).
Performance characteristics may differ in future versions.

\paragraph{Concurrency ceiling.}
The Phase~1 benchmark runs at \texttt{wrk -c30}, calibrated to keep
the five primary endpoints stable across the five main stacks.
The final primary dataset contains no \texttt{wrk}-reported errors
for those endpoints; instability observed during pilot calibration
therefore affects protocol selection, not the analysed primary rows.
This is a deliberately conservative choice: it favours apples-to-apples comparison of sustainable
throughput over exposing each stack's individual saturation behaviour.
A consequence is that runtime-level differences whose mechanisms
appear primarily under heavier concurrency (goroutine scheduling,
tokio work-stealing, V8 event-loop saturation, CPython+GIL contention,
.NET thread-pool growth) are not fully exercised at $c=30$.
Phase~2 addresses this limitation with a saturation sweep across
$c=30, 60, 120, 240$ on the \texttt{GET /items/:id} endpoint,
revealing divergent scaling: .NET scales linearly to $c=240$,
Go and Rust plateau near $c=120$, Node.js saturates at $c=120$,
and Python shows no meaningful scaling beyond $c=30$.
The sweep is limited to a single endpoint; saturation behaviour on
write-heavy endpoints may differ due to SQLite write-lock contention.

\paragraph{Multi-worker Python.}
Python was evaluated in Phase~1 with both a 4-worker granian
configuration (the primary comparison, denoted \texttt{python}) and a
single-worker variant (denoted \texttt{python-1w}).
The 4-worker deployment distributes requests across four OS
processes, achieving $\approx$1.3$\times$ higher throughput on read
endpoints at the cost of $\approx$2.6$\times$ higher RAM
(235\,MB vs.\ 89\,MB). The single-worker variant achieves
consistently better RPS/MB efficiency across all endpoints except
\texttt{get\_list} (where the I/O-bound serialisation of 1{,}000
rows benefits from multi-process parallelism).
This trade-off is particularly relevant in the self-hosted context,
where RAM is the binding constraint.

\paragraph{Power measurement accuracy.}
Power consumption was measured via the Raspberry~Pi~5's onboard PMIC (Dialog/Renesas DA9091)
through \texttt{vcgencmd pmic\_read\_adc}, the diagnostic interface documented by the
Raspberry~Pi Foundation. Absolute accuracy is limited to approximately
$\pm$10\,\% per the PMIC datasheet, and the sampling rate is bounded by the
\texttt{vcgencmd} interface to approximately 1\,Hz, which precludes analysis of
sub-second transients.
Because the four compiled/JIT stacks draw similar weighted average power
(3.72--4.27\,W), the $\pm$10\,\% uncertainty band ($\approx 0.4$\,W)
overlaps with part of the observed differences, making small energy
rankings between these stacks instrument-limited; only the Python gap
(5.57\,W weighted) sits outside the cluster.
This is the primary reason RQ4 is framed as exploratory rather than
confirmatory. We chose this method over an inline USB power meter
(e.g.\ UM25C) because such instruments introduce a measurable voltage drop
($\approx 0.13$\,V observed in our setup) that interferes with the Pi~5's PD
negotiation and can trigger undervoltage during sustained load, invalidating
the measurement. The PMIC measures the actual energy consumed by the SoC, RAM,
and on-board peripherals after the input regulator, which excludes regulator
losses but is invariant across language stacks. Since all stacks are
measured using the same instrument under identical conditions, relative
comparisons between stacks remain valid within the stated uncertainty.
Replication on dedicated external power profilers (e.g.\ Joulescope, Otii Arc)
is left for future work; these instruments were not available for this study
due to budget constraints.

\subsection{Construct Validity}

\paragraph{Throughput-to-RAM ratio as efficiency proxy.}
We operationalize ``efficiency'' as \texttt{rps\_per\_mb}
(throughput divided by peak~RSS). This single scalar metric may not
capture all dimensions of deployment efficiency (e.g., it ignores
startup time, binary size, and developer productivity).
We mitigate mono-operation bias by reporting nine metrics across four
research questions rather than relying on a single composite score.

\paragraph{Binary size heterogeneity.}
Binary size comparisons are not equivalent artifacts:
Go and Rust produce self-contained binaries;
Python and Node.js require an interpreter plus a dependency tree.
.NET Native AOT produces a self-contained executable.
We report both the entry artifact size and total deployment size
in \Cref{tab:rq3} and discuss the heterogeneity openly.

\paragraph{Energy measurement proxy.}
The PMIC-based energy measurement captures board-level SoC power,
not process-level energy. This includes power consumed by all
PMIC-monitored rails, not solely the CPU cores running the server.
Since all stacks are measured under identical conditions and the
server is the only active user-space process, relative comparisons
remain valid, but absolute energy-per-request values should not be
compared to measurements from different instruments.

\paragraph{CPU utilization ceiling.}
The \texttt{cpu\_avg\_pct} metric is process-scoped and not normalized
per core: on the Pi~5's four-core Cortex-A76, a single-threaded server
saturating one core reads $\approx$25\%, while a multi-threaded server
may reach $\approx$100\%. This makes cross-language comparisons
unintuitive for stacks with different threading models. We therefore
use CPU utilization only as a secondary diagnostic and do not include
it in the weighted aggregate or hypothesis tests.

\paragraph{Native server component (Python/granian).}
The Python stack uses granian, a Rust-implemented ASGI server, for
HTTP accept, parse, and dispatch. This means the measured Python
result includes native Rust code in the I/O hot-path and should be
interpreted as ``FastAPI deployment stack'' performance rather than
pure-Python runtime performance. We retain granian because (a)~it
preserves the ASGI application interface---all handler, serialization,
and database code remains Python---and (b)~it represents the dominant
deployment configuration for FastAPI in the self-hosted ecosystem as
of 2026. Nevertheless, the native server component inflates Python's
throughput relative to a pure-Python server such as uvicorn. This is
an asymmetry with Go (\texttt{net/http}, pure~Go) and Node.js
(Fastify, pure~JavaScript), where the HTTP server is written in the
same language as the application code. To bound the effect, we note
that the single-worker Python variant (\texttt{python-1w}) still
achieves the lowest throughput among all stacks despite the Rust
server component, suggesting that the CPython interpreter remains the
dominant bottleneck for request handling. A replication substituting
uvicorn for granian would quantify the exact contribution of the
native server and is left as future work.

\paragraph{Transpiled SQLite driver (Go/modernc).}
The Go stack uses \texttt{modernc.org/sqlite}, a mechanical
transpilation of the SQLite~C source into pure~Go via the
\texttt{ccgo} tool chain. This avoids CGo and enables trivial
cross-compilation to \texttt{linux/arm64}, but the transpiled code
does not benefit from the same compiler optimisations (vectorisation,
inlining, register allocation) that GCC or Clang apply to the
original~C source. Published micro-benchmarks report
\texttt{modernc.org/sqlite} at 10--40\% lower throughput than the
CGo-based \texttt{mattn/go-sqlite3} driver on write-heavy workloads.
All other stacks (Rust, Python, Node.js, .NET) link against the
natively compiled~C SQLite library. This creates an asymmetry on
database-bound endpoints where SQLite execution time dominates: Go's
measured throughput may understate what the Go runtime could achieve
with a native~C SQLite backend. The bias direction is \emph{against}
the Go-centred hypotheses (H1, H2), making the reported conclusions
more conservative. We retained \texttt{modernc.org/sqlite} because
reproducibility without CGo was a design priority; a sensitivity
replication using \texttt{mattn/go-sqlite3} would bound the effect.

\subsection{Conclusion Validity}

\paragraph{Multiple comparisons.}
Across five endpoints and eight endpoint-variant metrics, each with
$\binom{5}{2}=10$ pairwise Dunn comparisons, we conduct 400 post-hoc
tests. Individual result tables report Bonferroni correction
within each metric (10 comparisons). As a robustness check, we also
applied Holm-Bonferroni step-down correction across the full 400-test
family: 253 of 400 comparisons (63.25\%) remain significant at
$\alpha=0.05$. The 147 tests that lose significance are predominantly
between similarly-performing compiled/JIT stacks (e.g.\ Go vs.\ Rust
RPS, .NET vs.\ Node.js power) where per-metric Bonferroni $p$-values
were borderline ($0.0004$--$0.05$) and Cohen's $d$ values are small.
All comparisons involving large effect sizes ($d > 0.8$) survive
the family-wide correction. The full Holm-corrected results are
included in the replication package
(\texttt{holm\_bonferroni\_family.csv}).

\paragraph{Endpoint-type variance (resolved).}
When throughput is pooled across five heterogeneous endpoints ($N=300$
per language), endpoint type explains most of the variance (Kruskal-Wallis
$\eta^2_H = 0.061$ for language effect on pooled RPS).
We resolve this confound by analyzing each endpoint independently
($N=50$ per language per endpoint), which reveals large language effects
on all endpoints ($\eta^2_H = 0.63$--$0.96$).

\paragraph{Workload-mix assumption.}
The weighted aggregate metrics depend on the declared read-heavy
workload mix (30/35/15/10/10 for GET-list / GET-one / POST / PUT / DELETE).
This mix is a modeling choice, not an empirical observation.
Different applications will have different traffic distributions, and
the weighted rankings could change under alternative mixes.
We mitigate this threat with a sensitivity analysis across three mixes:
\Cref{tab:sensitivity} shows that the language ranking for weighted
throughput is stable across read-heavy, uniform, and write-heavy mixes.
The four compiled/JIT stacks maintain overlapping CIs under all three mixes,
and Python remains significantly below the cluster in every scenario.

\begin{table}[ht]
\centering
\small
\caption{Sensitivity analysis: weighted throughput (req/s) under three
         workload mixes (5 core endpoints).
         Rankings are stable across all mixes.}
\label{tab:sensitivity}
\begin{tabular}{lrrr}
\toprule
\textbf{Language} & \textbf{Read-heavy} & \textbf{Uniform} & \textbf{Write-heavy} \\
\midrule
.NET    & 2{,}461 & 2{,}240 & 2{,}350 \\
Node.js & 2{,}254 & 2{,}084 & 2{,}252 \\
Go      & 2{,}200 & 2{,}088 & 2{,}295 \\
Rust    & 2{,}125 & 2{,}013 & 2{,}219 \\
Python  &   969   &   887   &   964 \\
\bottomrule
\end{tabular}
\end{table}

% ============================================================
\section{Conclusion}
\label{sec:conclusion}
% ============================================================

We presented a controlled experiment measuring five language stacks
(Go, Rust, Python/Granian, Node.js/Fastify, .NET~10 Native AOT) on a
Raspberry~Pi~5, focusing on the metrics most relevant to self-hosted
software deployment: RAM footprint, throughput-to-RAM ratio, startup
time, and energy consumption.
A three-phase protocol (sustained-load profiling at $N=50$ runs per
language per endpoint; concurrency saturation sweep at $c=30$--$240$;
heap-memory time-series) resolves endpoint-type confounds, quantifies
scaling behaviour, and characterises runtime memory management
strategies.

Our main findings, all framed as pilot observations on a single
Raspberry~Pi~5 unit under a SQLite-backed CRUD workload, are:
\begin{enumerate}[nosep]
  \item \textbf{Throughput-to-RAM ratio.}
    Rust shows the highest weighted RPS/MB
    (310.25\,req/s/MB under the read-heavy mix), $2.8\times$ higher
    than Go (111.13) and $16.4\times$ higher than Node.js (18.93).
    .NET leads weighted raw throughput (2{,}461\,req/s), followed by
    Node.js (2{,}254), Go (2{,}200), and Rust (2{,}125); all four
    compiled/JIT stacks form a throughput cluster with overlapping
    bootstrap CIs. Only Python/4-worker Granian is consistently
    slower (969\,req/s).
  \item \textbf{RAM footprint.}
    Weighted peak RAM spans a $32.2\times$ range: from 7.36\,MB
    (Rust) to 236.71\,MB (Python). Heap-snapshot time-series
    (Phase~3) reveal that Node.js grows monotonically without visible
    GC cycles (peak 233.8\,MB, $+117$\,MB growth), while Rust holds
    a flat 7.8\,MB with zero GC events, and Go exhibits 36 visible
    GC pauses while remaining below 20\,MB.
  \item \textbf{Concurrency scaling.}
    Under the saturation sweep (Phase~2, \texttt{GET /items/:id}),
    .NET scales linearly to $c=240$ (26{,}209\,req/s), Rust and Go
    plateau near $c=120$ (17{,}469 and 14{,}230\,req/s at $c=240$),
    Node.js saturates at $c=120$ (9{,}766\,req/s at $c=240$), and
    Python shows negligible scaling (2{,}107\,req/s at $c=240$).
  \item \textbf{Energy per request (exploratory).}
    Energy distributions are heavily right-skewed; we report
    weighted medians with bootstrap CIs.
    The four compiled/JIT stacks fall inside a 2.71--5.88\,mJ band.
    Because the ranking is partly driven by throughput and measured
    with a ${\pm}10\%$ PMIC at 1\,Hz, we do not interpret the
    within-cluster ordering as a confirmatory energy finding.
    Python (17.42\,mJ; 5.57\,W weighted average) is the only stack
    outside the measurement uncertainty.
    Confirmatory energy comparisons require a calibrated external
    power profiler.
  \item \textbf{.NET~10 Native AOT.}
    Starts in 44\,ms, weighted peak RAM of 47.65\,MB, and draws
    comparable average power to Rust and Go ($\approx$3.1\,W) on this
    Pi~5 unit. Under the saturation sweep, .NET is the only stack
    that continues scaling linearly at $c=240$, suggesting it is a
    strong candidate for self-hosted Pi~5 deployment, particularly
    for read-heavy workloads.
  \item \textbf{Go hypothesis.}
    H$_1$ (best throughput-to-RAM ratio) is not supported by these
    pilot data: Rust exceeds Go on RAM footprint ($2.7\times$ lower)
    and on the RPS/MB ratio ($2.8\times$ higher weighted).
    H$_2$ (competitive raw throughput and startup) is consistent with
    the data: the four compiled/JIT stacks show overlapping throughput
    CIs.
  \item \textbf{Python workers trade-off.}
    The single-worker Python variant (python-1w) achieves higher
    RPS/MB efficiency than the 4-worker deployment on 4 of 5 core
    endpoints, at the cost of lower absolute throughput.
    This demonstrates that right-sizing worker count is critical on
    RAM-constrained devices.
\end{enumerate}

Per-endpoint analysis reveals that all five core endpoints now show
significant language effects (Kruskal-Wallis $p < 0.001$ for all),
with $\eta^2_H$ ranging from 0.63 (\texttt{post}) to 0.96
(\texttt{get\_list}). The \texttt{get\_list} endpoint, which was
non-significant in a prior N=30 pilot, emerges as the most strongly
differentiated endpoint: .NET leads at 1{,}301\,req/s vs.\
Go at 392\,req/s, likely due to system-level serialisation
efficiency when returning 1{,}000-row JSON payloads.
Throughput convergence on write endpoints (\texttt{post},
\texttt{put}, \texttt{delete}) is partly an artefact of SQLite's
file-level write lock, which caps the achievable rate independent of
the runtime; we discuss this ceiling explicitly in
\Cref{sec:threats}. Sensitivity analysis across three workload mixes
(\Cref{tab:sensitivity}) confirms that the qualitative language
ordering is stable across read-heavy, uniform, and write-heavy
distributions.

The selected concurrency level ($c=30$), together with endpoint-specific
database seeding, produced a clean primary dataset for the five core
endpoints and five main stacks.
The practical implication, under the scope of this pilot
(single Pi~5 unit, CRUD-over-SQLite workload), is that the data are
consistent with Rust offering the best measured resource efficiency
when RAM is the binding constraint, and with .NET offering the best
absolute throughput and concurrency scaling. Go remains a practical
alternative when ecosystem breadth and developer ergonomics (not
measured) are prioritised.
We caution against extrapolating these observations to other ARM64
boards, to non-SQLite storage backends, or to workloads with
different read/write ratios without replication.

Future work includes: replication on additional ARM64 boards
(Raspberry~Pi~4, Orange~Pi~5, Rock~5) and on additional units of
the Pi~5 to bound unit-to-unit variance; replacement of SQLite with a
client-server database (e.g.\ PostgreSQL) to lift the write-lock
ceiling and isolate runtime effects from storage contention; energy
measurement with a calibrated external power profiler (Joulescope,
Otii Arc) to move RQ4 from exploratory to confirmatory; and extension
of the concurrency sweep to write-heavy endpoints where SQLite lock
contention may produce different saturation profiles.

% ============================================================
% Data Availability
% ============================================================

\section*{Data Availability}

\ifanonymized
  The replication package is available in an anonymous repository.
  The full URL will be disclosed upon acceptance.
\else
  The replication package, including raw benchmark data, analysis scripts,
  server implementations, and plotting scripts, is available at:

  \url{https://github.com/woliveiras/2026-lang-self-hosted-pi}.
\fi
The Phase~1 dataset file, \path{benchmarks_20260520T125750Z.csv},
has SHA-256 hash \path{839362eb...3d07482}.\footnote{Full hash:
\texttt{839362ebbf2a17e4e4666bba1f5deacfdcf68a85b2486e53e322610413d07482}.}

The Phase~2 sweep data file, \path{benchmarks_20260521T071647Z.csv},
has SHA-256 hash \path{f6162325...dded5c7c}.\footnote{Full hash:
\texttt{f6162325dc287176b5c5a1b996c040836dd8a963d5344e814b569929dded5c7c}.}

Phase~3 memory time-series data contain 3{,}314 JSON snapshots.
They are stored in \path{data/heap-snapshots/}, one file per run,
with timestamped RSS readings at $\approx$130\,ms resolution.
The firmware, bootloader, and PMIC metadata captured on the tested
Raspberry~Pi~5 unit at data-collection time are recorded in
\path{data/firmware_info.txt} in the replication package.

A reproducibility container (\path{analysis/Dockerfile}) regenerates
all tables and plots in this paper from the published CSV; see
\path{analysis/README.md} for the build and run commands.

% ============================================================
% Bibliography
% ============================================================

\bibliographystyle{plain}
\bibliography{references}

\end{document}